\documentclass[english, 10pt]{article}
\usepackage{authblk}
\usepackage{multicol}
\usepackage[a4paper]{geometry}
\usepackage[utf8]{inputenc}
\usepackage[T1]{fontenc}
\usepackage{babel}
\usepackage{fancyhdr}
\usepackage{graphicx}
\usepackage{amsmath,amsfonts,amssymb}
\usepackage{booktabs}
\usepackage[T1]{fontenc}
\usepackage{listings}
\usepackage{color}
\usepackage{hyperref}
\usepackage{lineno}
\usepackage{csquotes}
\hypersetup{colorlinks=true, linkcolor=blue, urlcolor=blue, citecolor=blue, pdfborder={0 0 255}}
\usepackage[natbib=true,style=numeric,sorting=none]{biblatex}
\DeclareUnicodeCharacter{0301}{\'{e}}
\usepackage{colortbl}
\usepackage{url}
\usepackage{caption}
\usepackage[toc, page]{appendix}
\usepackage{subcaption}
\usepackage{dirtytalk}
\usepackage{lineno}
\usepackage{setspace}
\usepackage{lipsum}
\usepackage{bm}
\usepackage[ruled,vlined]{algorithm2e}
\usepackage{amssymb}
\usepackage{amsmath,amsthm}

\usepackage{mathtools}
\usepackage[thinc]{esdiff}

\usepackage{listings}
\usepackage{xcolor}

\definecolor{codegreen}{rgb}{0,0.6,0}
\definecolor{codegray}{rgb}{0.5,0.5,0.5}
\definecolor{codepurple}{rgb}{0.58,0,0.82}
\definecolor{backcolour}{rgb}{0.95,0.95,0.92}

\lstdefinestyle{mystyle}{
    backgroundcolor=\color{backcolour},   
    commentstyle=\color{codegreen},
    keywordstyle=\color{magenta},
    numberstyle=\tiny\color{codegray},
    stringstyle=\color{codepurple},
    basicstyle=\ttfamily\footnotesize,
    breakatwhitespace=false,         
    breaklines=true,                 
    captionpos=b,                    
    keepspaces=true,                 
    numbers=left,                    
    numbersep=5pt,                  
    showspaces=false,                
    showstringspaces=false,
    showtabs=false,                  
    tabsize=2
}

\title{\textbf{Kinematics Modeling of Peroxy Free Radicals: A Deep Reinforcement Learning Approach}}
\date{}

\author[1]{Subhadarsi Nayak\thanks{Corresponding Author:- Subhadarsi Nayak - \href{mailto: subhadarsinayak1992@gmail.com}{\texttt{subhadarsinayak1992@gmail.com}}}}
\author[2]{Hrithwik Shalu}
\author[3]{Joseph Stember\thanks{Senior Author}}

\affil[1]{Department of Chemistry, Indian Institute of Technology Madras}
\affil[3]{Department of Aerospace Engineering, Indian Institute of Technology Madras}
\affil[4]{Memorial Sloan Kettering Cancer Center, New York, NY, US, 10065}

\makeatletter
\let\runtitle\@title
\runtitle

\begin{document}

\maketitle
\begin{abstract}

Tropospheric ozone, known as a concerning air pollutant, has been associated with health issues including asthma, bronchitis, and impaired lung function. The rates at which peroxy radicals react with NO play a critical role in the overall formation and depletion of tropospheric ozone. However, obtaining comprehensive kinetic data for these reactions remains challenging. Traditional approaches to determine rate constants are costly and technically intricate. Fortunately, the emergence of machine learning-based models offers a less resource and time-intensive alternative for acquiring kinetics information. In this study, we leveraged deep reinforcement learning to predict ranges of rate constants (\textit{k}) with exceptional accuracy, achieving a testing set accuracy of 100\%. To analyze reactivity trends based on the molecular structure of peroxy radicals, we employed 51 global descriptors as input parameters. These descriptors were derived from optimized minimum energy geometries of peroxy radicals using the quantum composite G3B3 method. Through the application of Integrated Gradients (IGs), we gained valuable insights into the significance of the various descriptors in relation to reaction rates. We successfully validated and contextualized our findings by conducting cross-comparisons with established trends in the existing literature. These results establish a solid foundation for pioneering advancements in chemistry, where computer analysis serves as an inspirational source driving innovation.

\end{abstract}

\section*{Introduction}

In the field of atmospheric studies, there has been a great deal of interest in global tropospheric chemistry and its influence on climate~\cite{duce1983organic, calvert1985chemical, ainsworth2012effects}.
In general, the troposphere is considered the region of the atmosphere situated 10-18 km from the earth's surface, depending on the latitude and season. The temperature ranges from 210-289 K (Kelvin), and pressure averages 1013 mb (millibar) at Earth's surface to 140 mb near the tropopause region~\cite{highwood1998tropical}. The abundance of tropospheric ozone is changing significantly; photochemical pollution plays a crucial role in ozone $(\mathrm{O_3})$ formation in the tropospheric region. The main sources of hazardous chemical compounds that are emitted into the troposphere as a result of human activities can broadly be classified as anthropogenic (pollution)~\cite{berntsen1997effects} and biogenic sources~\cite{ganzeveld2002global}. From the existing literature, it is evident that anthropogenic NO\textsubscript{x} emissions cause high ozone concentrations~\cite{liu1987ozone, wild2001indirect, zhang2003impacts, lamsal2011application}. 

One point of potential confusion regarding the ozone $(\mathrm{O_3})$ layer is that in the stratosphere it plays an important and actually \textit{desired} role in restricting the transmission of ultraviolet radiation below 290 nm, noting that the latter contributes to skin cancer. In contrast to the stratosphere, in the lower troposphere, high $(\mathrm{O_3})$ levels are associated with \textit{adverse} effects on human health; concentrations of more than 70 ppb (parts per billion) in the are considered to be hazardous~\cite{crutzen1988tropospheric} and are associated with health issues including asthma, bronchitis, and impaired lung function. Hence it can be stated that in the troposphere as opposed to within the stratosphere, high ozone $(\mathrm{O_3})$ concentration makes it a \textit{dangerous} air pollutant that has serious effects on the ecosystem~\cite{ainsworth2012effects}. 

Numerous experimental research has been conducted with the goal of identifying the danger, important biological effects, and doses pertinent to health declines following trophospheric ozone exposure~\cite{witschi1999ozone, paige1999acute}. Globally, anthropogenic $\mathrm{O_3}$ pollution is thought to be responsible for 0.7 million annual fatalities~\cite{jaffe1967biological, ainsworth2012effects, iriti2008oxidative}. In addition to lung damage in humans, crop yields are negatively impacted by ozone, with negative effects on stomatal conductance, photosynthetic carbon assimilation, and plant growth~\cite{ainsworth2012effects}. 

One of the main sources of tropospheric $\mathrm{O_3}$ formation is via the photolysis of $\mathrm{NO_2}$~\cite{wang2018attribution}, as shown in the following pair of reactions:

\begin{equation}
    \text{NO}_2 + h \nu \xrightarrow{} \text{NO} + \text{O}(^{3}\text{P})
\label{chem_eq_1}
\end{equation}

\begin{equation}
    \text{O}(^{3}\text{P}) + \text{O}_2 + \text{M} \xrightarrow{} \text{O}_{3} + \text{M}
\label{chem_eq_2}
\end{equation}

\noindent
It has been found that a significant source of atmospheric reactive carbon, and ultimately the reactant $\text{NO}_2$ in Reaction \eqref{chem_eq_1}, is vegetation~\cite{heiden2003emissions}. Most reactive carbon emissions from plants are currently understood to comprise volatile olefinic substances such as isoprene and monoterpenes~\cite{fehsenfeld1992emissions, pinto2010plant}. The oxidation reaction of such volatile organic compounds (VOCs) leads to the formation of the intermediate $\dot{\text{RO}_2}$ radical, which plays a key role in the formation of $\mathrm{NO_2}$ in the troposphere. In particular, $\dot{\text{RO}_2}$ is a reactant in the free radical reaction that converts existing {NO} in the troposphere to $\mathrm{NO_2}$:

\begin{equation}
    \dot{\text{RO}_2} + \text{NO}  \xrightarrow{} \dot{\text{RO}} + \text{NO}_2
\label{chem_eq_3}
\end{equation}

\noindent
Studying the factors contributing to ozone $\mathrm{O_3}$ depletion in the troposphere is important in understanding the balance with ozone production. The photolysis of $\mathrm{O_3}$ leads to the formation of $\text{O}(^{1}\text{D})$ which, followed by reaction with water vapor, forms $\dot{\text{OH}}$ free radicals, the same results in a net loss of tropospheric ozone~$\mathrm{O_3}$. In the setting of sufficiently low concentration of NO, $\mathrm{O_3}$ reacts with $\dot{\text{OH}}$ and $\dot{\text{HO}_2}$ radicals. The corresponding reactions of ozone~$\mathrm{O_3}$ with these two free radical reactions are an additional source for the loss of tropospheric ozone. The concentration of NO plays an important role in the net formation and loss of $\mathrm{O_3}$ in the troposphere. The net formation of $\mathrm{O_3}$ is determined by the rate of reaction of $\mathrm{RO_2}$ radical with NO as given by the Reaction~\ref{chem_eq_3}. The number of NO molecules converted to $\mathrm{NO_2}$ plays a key role in deciding the rate of formation of $\mathrm{O_3}$ in troposphere~\cite{atkinson2000atmospheric}. Because of the radioactive properties of $\mathrm{O_3}$ in the ultraviolet and the infrared wavelength range, it plays an essential role in deciding the radiation budget of the earth's atmosphere ~\cite{oyama2000chemical}. This characteristic absorption of ultraviolet radiation protects the biosphere from harmful radiation~\cite{crutzen1979role}. Considering the crucial nature of free radical reactions in their contribution for determining ozone $\mathrm{O_3}$ dynamics in the atmosphere, we aim to gain further insights into the rate contribution factors associated with the nature of hydrocarbons that undergo such reactions.\\

\noindent
In recent years machine learning (ML)-based approaches have increasingly replaced classical methods to study reaction kinetics and mechanisms. The adoption is due largely to ML's better prediction accuracies and computational cost-savings ~\cite{meuwly2021machine, chen2022machine, stocker2020machine}. A recent study~\cite{beker2019prediction}, for example, showed that ML-based prediction of key descriptors/parameters of reaction conditions that determined kinetics could achieve higher accuracy compared to standard quantum mechanical (QM) methods. ML also saved computational time -- around 0.5 seconds versus several hours for standard QM-based predictions ~\cite{beker2019prediction}. In the context of synthetic organic chemistry ML techniques have been used for the prediction of feasible reaction paths and products~\cite{singh2020unified}. From existing literature, it can be seen that ML-based methods play a significant role in understanding atmospheric reactions~\cite{meuwly2021machine}. In~\cite{jorner2021machine}, a Gaussian Process Regression model was used for the purposes of deciding the activation energies of nucleophilic aromatic substitution reactions. Free radical reaction pathways were exhaustively investigated in~\cite{lee2020graph} without requiring chemical intuition, using a graph theory-based reaction pathway search approach. In~\cite{sanches2022evaluating}, an ML-based approach was used to model the mechanism and kinetics of OH radical reactions.  ML showed greater predictive capacity than traditional alternatives. 

It is important to note that kinematics modeling based on traditional ML approaches requires substantial domain expertise, as poor decisions of feature descriptors of the chemical system under consideration at any step will lead to large modeling errors. A notable alternative is to use Deep Learning~\cite{lecun2015deep} based approaches that form self-adapting descriptors that represent meaningful information from raw data. A recent work~\cite{mardt2018vampnets} developed a deep learning framework for molecular kinetics using neural networks that provided easily interpretable few-state kinetic models as compared to traditional Markov models. A framework for screening candidate reaction solvents was devised in~\cite{wu2022reaction}, and chemical descriptors based on the frontier molecular orbital theory were combined with a deep learning-based reaction kinetics model. In~\cite{kollenz2020unravelling}, a deep learning-based framework was developed in an effort to obtain the kinetic model of photochemical reactions. Deep learning was more robust in terms of experimental noise and typical pre-analysis errors like time-zero corrections. In \cite{zhong2020deep}, a Deep Neural Network (DNN) with molecular fingerprints as feature descriptors was developed to directly predict the $\dot{\text{OH}}$ radical rate constants with respect to 593 organic contaminants. An ensemble model combining the DNN with XGBoost was introduced in \cite{zhong2021shedding} for estimating the reactivity of $\dot{\text{OH}}$ radicals toward newly emerged organic compounds. 

Building upon the existing literature, in this work, we sought to design a practical ML-based approach in order to estimate the nature of kinetics of reactions involving various VOCs, the mechanism shown in Reaction \ref{chem_eq_3}. A significant challenge faced in the effective implementation of such a method is that for the particular class of reactions under consideration, there are very few experimental and computational studies reported in the existing literature. We compare the relative performances of various ML-based approaches, including a recently introduced Reinforcement Learning (RL)-based data efficient alternative~\cite{stember2022deep} to determine the optimality of the kinetics modeling obtained. A much broader goal here is to allow the kinematics model correlations with respect to input physical descriptors to provide a source of inspiration to explore new insights involved in the kinetics of the reaction of peroxy radicals with Nitric Oxide~(NO).
\pagebreak

\section{Methods}

In this section, we outline the set of methods used to analyze chemical kinematics data. The kinematics modeling used here is either a simplified standard neural regression task (to predict the reaction rate constant) or a NN-based classification task for estimating the possible order of kinematics. The limitation of data for the task is acknowledged in this study; standard forms of conventional methods in the field of deep learning (DL) are tried out for a rigorous comparison.

\subsection{Neural Regression / Classification}

\begin{figure}[h!]
    \centering
    \includegraphics[width=14cm, height=7cm]{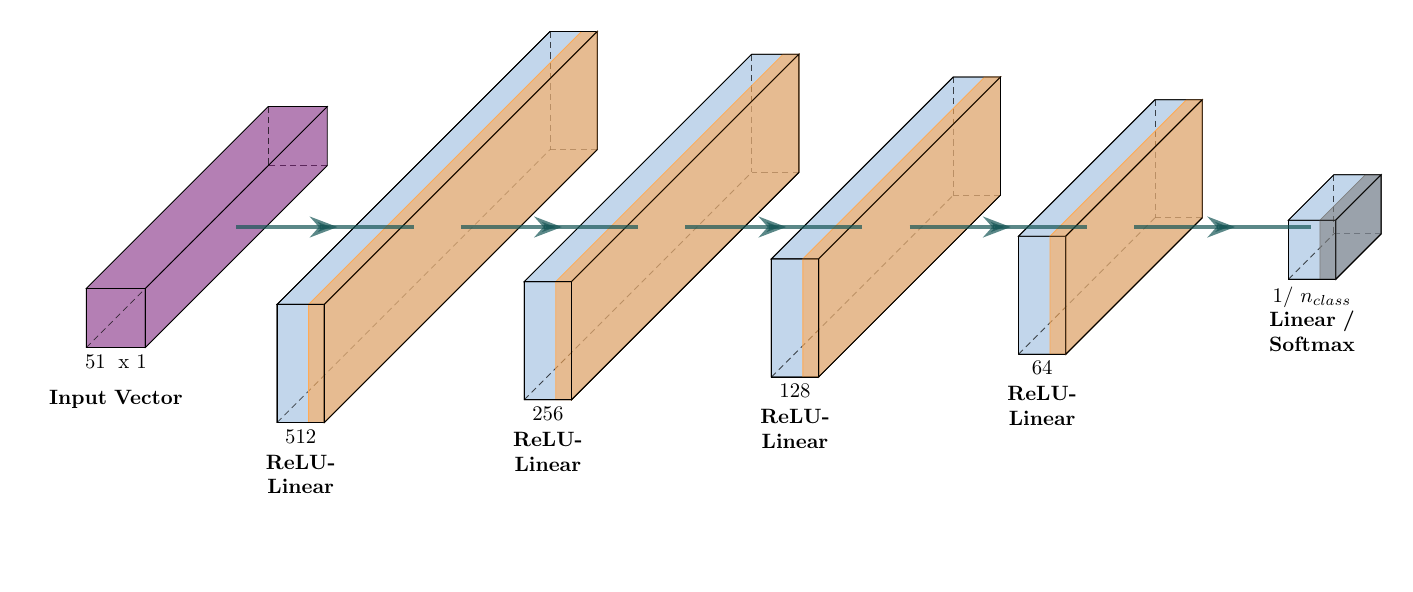}
    \caption{Architecture of the Neural Network used for continuous or discrete predictions.}
    \label{3_1}
\end{figure}

We used regression in its most standard form via an Artificial Neural Network (ANN) to model a set of inputs related to a continuous prediction. The target variable for regression/classification is selected is the rate constant, \textit{k}. We made use of a Neural Network Architecture whose hidden layer neuron scheme follows a monotonic exponential relation with the number of hidden layers. The ANN architecture used for this study is shown in Figure \ref{3_1}. We arrived upon the architecture after on a rigorous random search \cite{liashchynskyi2019grid} for hyperparameters. 

\noindent
To normalize the target variable, we used the standard min-max scaling procedure:

\begin{equation}
    y_{scaled} = \frac{y - y_{min}}{y_{max}-y_{min}}
\label{eq1}
\end{equation}

\noindent
Note that, apart from the normalization, a log scale transformation \cite{yeo2000new} was applied to the target variable apriori so as to improve its distribution. The objective function used for the purposes of ANN-based regression was the standard mean squared error, which in its standard form (applied to $N$ data points) is as follows:

\begin{equation}
    MSE\;(y,\; y_{pred}) = \sum^N_{i=1} \left| y^i - y^i_{pred} \right|^2
\label{eq2}
\end{equation}

\noindent
For the purpose of obtaining a discretized set of labels for classification, we performed a multi-step clustering process as outlined in Algorithm \ref{algo1}. To obtain an optimal and reproducible set of clusters, we used the k-means++ method \cite{arthur2006k} for all clustering steps mentioned in Algorithm \ref{algo1}.

\begin{algorithm}[!h]
\SetAlgoLined
\KwData{ Target Variable - {$y$}}
\KwResult{Obtain $Y$ (Class labels)}

\textbf{Obtain initial set of clusters and centroids}\\
$Y_i$, $C_i$ = K-means++ ($y$)\;

\textbf{Select majority class from the set of clusters}\\
$y_m$, $c_m$ = $\max$ $(N(Y_i))$\;

 \textbf{Iterate until convergence is achieved}\\
 \While{$iter$ < $iter_{max}$}{
 
 \If{$class\; balanced$}{
 break\;
 }
 
 \textbf{Cluster on the majority class}\\
 $Y_j$, $C_j$ = K-means++ ($y$ = $y_m$)\;
 
 \textbf{Repopulate the set of clusters}\\
 $Y_i$  = $Y_i$ + $Y_j$\;
 
 \textbf{Re-select majority class from the new set of clusters}\\
 $y_m$, $c_m$ = $\max$ $(N(Y_i))$\;
 }
 \textbf{Return final set of clusters : $Y_i$}\\
 \caption{Multi-Step Clustering for Data Discretization}
 \label{algo1}
\end{algorithm}

\noindent
The objective function we used for the vanilla form of ANN based classification was cross-entropy, the standard form (applied to $N$ data points) as shown in Equation \ref{eq3}.

\begin{equation}
    CE\;(y,\; y_{pred}) = -\sum^N_{i=1}\sum^{N_{class}}_{j=1} y^{ij} \log (y^{ij}_{pred})
\label{eq3}
\end{equation}

\noindent
We used the Adam \cite{zhang2018improved} optimizer with a base learning rate of ($10^{-3}$) for all ANN training purposes. Note that all tunable hyperparameters used were obtained through random search.

\subsection{Classification by Representation}

In studies such as this, where there is an immense shortage of labeled data, learning a compact latent vector representation model for a given set of data (for effective comparison) is a suitable alternative, rather than learning a direct functional mapping from data to labels. Recent work \cite{khosla2020supervised} outlines the effectiveness of contrastive objectives for supervised representation learning in tasks where data is sparse, and label information is effectively utilized. For our purposes, we used the following form of the contrastive loss \cite{schroff2015facenet}:

\begin{equation}
    \mathcal{L}_{contrastive} (y_1,\; y_{2}) = (1 - B)\frac{1}{2}\left(\mathcal{D}(y_1,\; y_{2})\right)^2 + (B)\frac{1}{2} \left[\max\left(0, m - \mathcal{D}(y_1,\; y_{2})\right)\right]^2
\label{eq4}
\end{equation}

\noindent
Note that in Equation \ref{eq4}, the Boolean label $B$ signifies whether or not both latent representations $(y_1, y_2)$ have the same categorical label. We chose the distance metric $\mathcal{D}$ as the euclidean between the vectors. The margin parameter $m$ decides the bound on the magnitude of the overall contrastive objective, acting as an important hyperparameter for the overall setting.

\begin{figure}[h!]
    \centering
    \includegraphics[width=14cm, height=10cm]{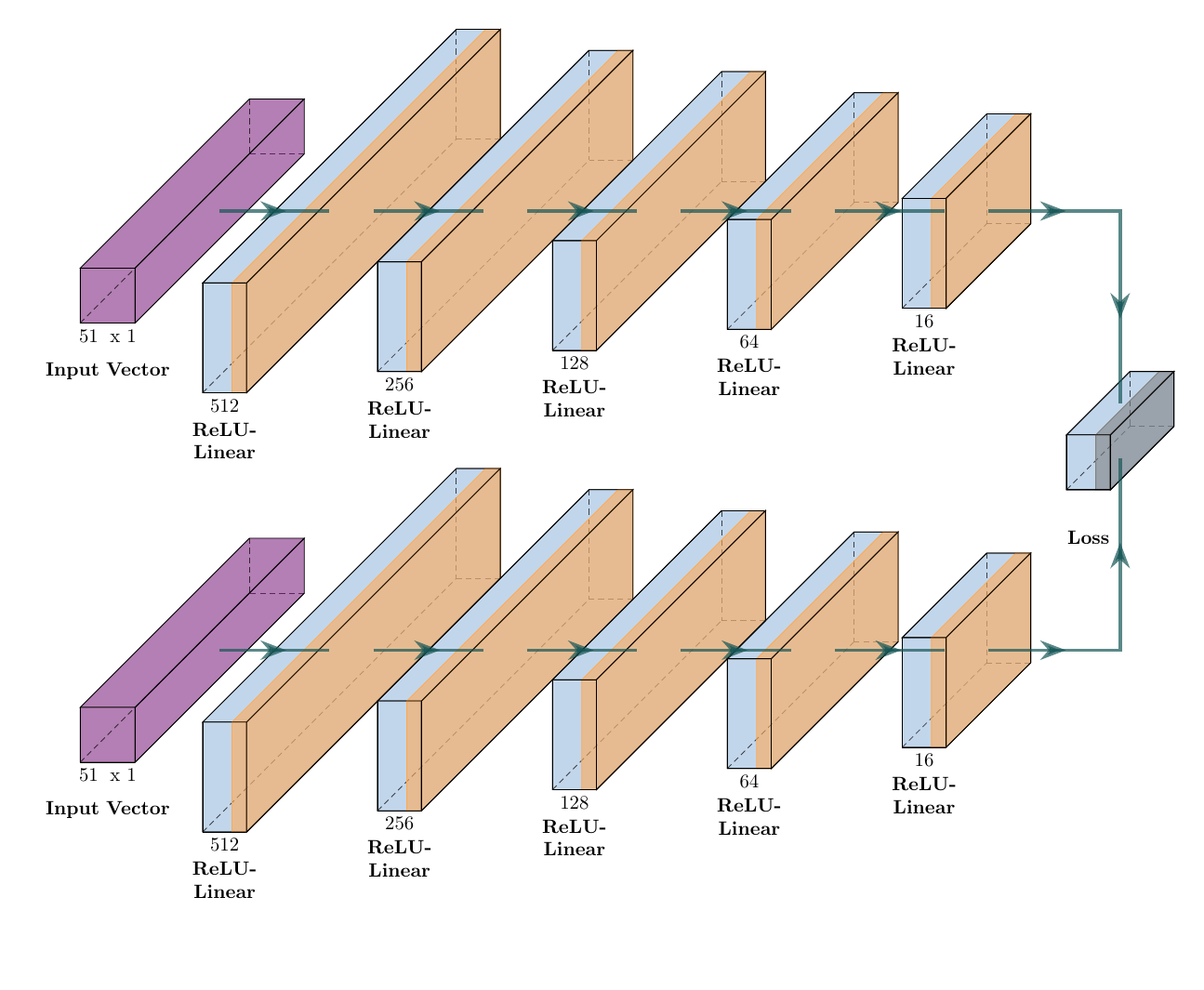}
    \caption{Siamese neural network architecture used in the representation learning framework. }
    \label{3_2}
\end{figure}

\noindent
In order to generate the pair of embeddings required for the framework at each step, as outlined in the objective (Equation \ref{eq4}) we made use of the Siamese network architecture (Shared Weights) as illustrated in Figure \ref{3_2}.

\subsection{Classification by Reinforcement Learning}


In this section, we outline the adaptation of a deep reinforcement learning (DRL)-based approach (as Introduced in \cite{stember2022deep}) for optimal classification in sparse data settings. The approach is primarily based on the Deep Q-Learning \cite{mnih2015human} method, where classification is reformulated as a multi-step decision-making process instead of single-step prediction modeling. The advantage of doing so is the additional control on the variance of the objective function over the training period. Doing so markedly improves the algorithm's overall data efficiency by allowing a wider search space for parameter optimization. As in any classic RL setting, the overall goal is to form an agent (policy) that performs optimal actions in given state, where the actions' optimality is governed by scalar feedbacks (rewards). More formally, Q-Learning \cite{watkins1992q} proceeds through the optimization of Q-value (action value), which for a given policy ($\pi$) and a discounted Markov Decision Process (MPD) is as outlined in Equation \ref{eq5}.  

\begin{equation}
Q^\pi(s_t, a_t) = \mathbf{E} \left [ R_{t + 1} + \gamma R_{t+2} + \gamma^2 R_{t+3} + ... \;| \;(s_t, a_t) \right]
\label{eq5}
\end{equation}

\noindent
Q-Learning proceeds by the principle of temporal difference learning, where the Q-value estimate of a given step (training epoch) is compared with an estimated target using the bellman optimality equation. We used the following modified form of the Bellman equation \cite{sutton1999reinforcement} to estimate the target Q-value:

\begin{equation}
Q^\pi_{target}(s_t, a_t) = R_{t + 1} + \gamma \max_a \{Q^\pi(s_{t+1}, a)\}
\label{eq6}
\end{equation}

\begin{figure}[h!]
    \centering
    \includegraphics[width=14cm, height=8cm]{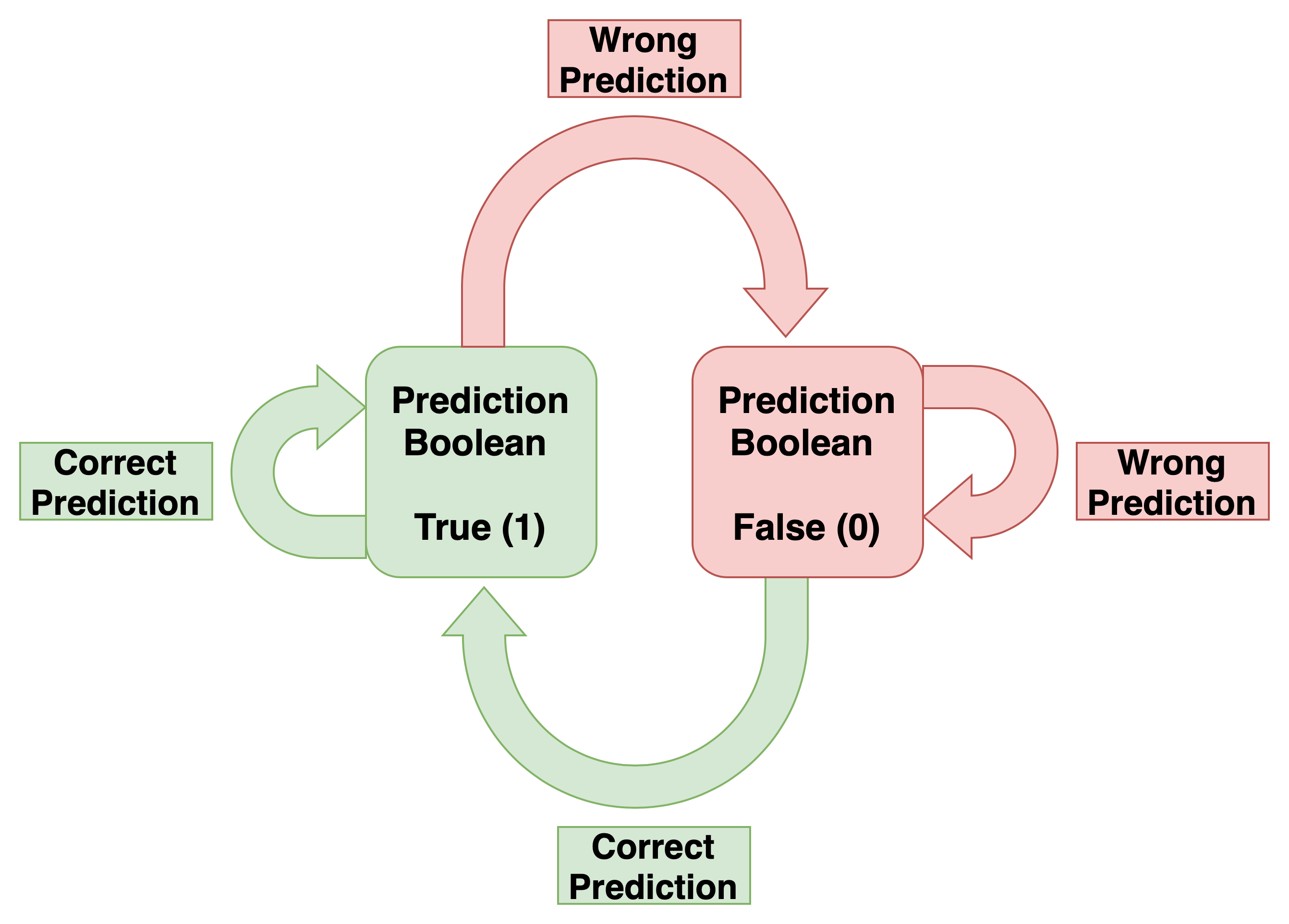}
    \caption{The simplistic Markov Decision Process used, illustrating the state-action interplay.}
    \label{3_3}
\end{figure}

\noindent
Note that we used a simple two-state deterministic Markov Decision Process (MDP) to model the state transition dynamics. In this formulation, a single state is defined by a data point and a Boolean that indicate the prediction's correctness. The MDP is formally illustrated in Figure \ref{3_3}.

\noindent
We used a binary reward that depended on the state-action pair, as shown in Equation \ref{eq7}. Note that the prediction Boolean can vary with the reward's nature.

\begin{equation} 
R_t =  \begin{cases}
        +1 & \text{if } a_{t-1} =  \text{label} 
        \\ -1 & \text{if } a_{t-1} \neq \text{label}
        \text{.}
    \end{cases}
\label{eq7}
\end{equation}

\begin{figure}[!ht]
    \centering
    \includegraphics[width=14cm, height=10cm]{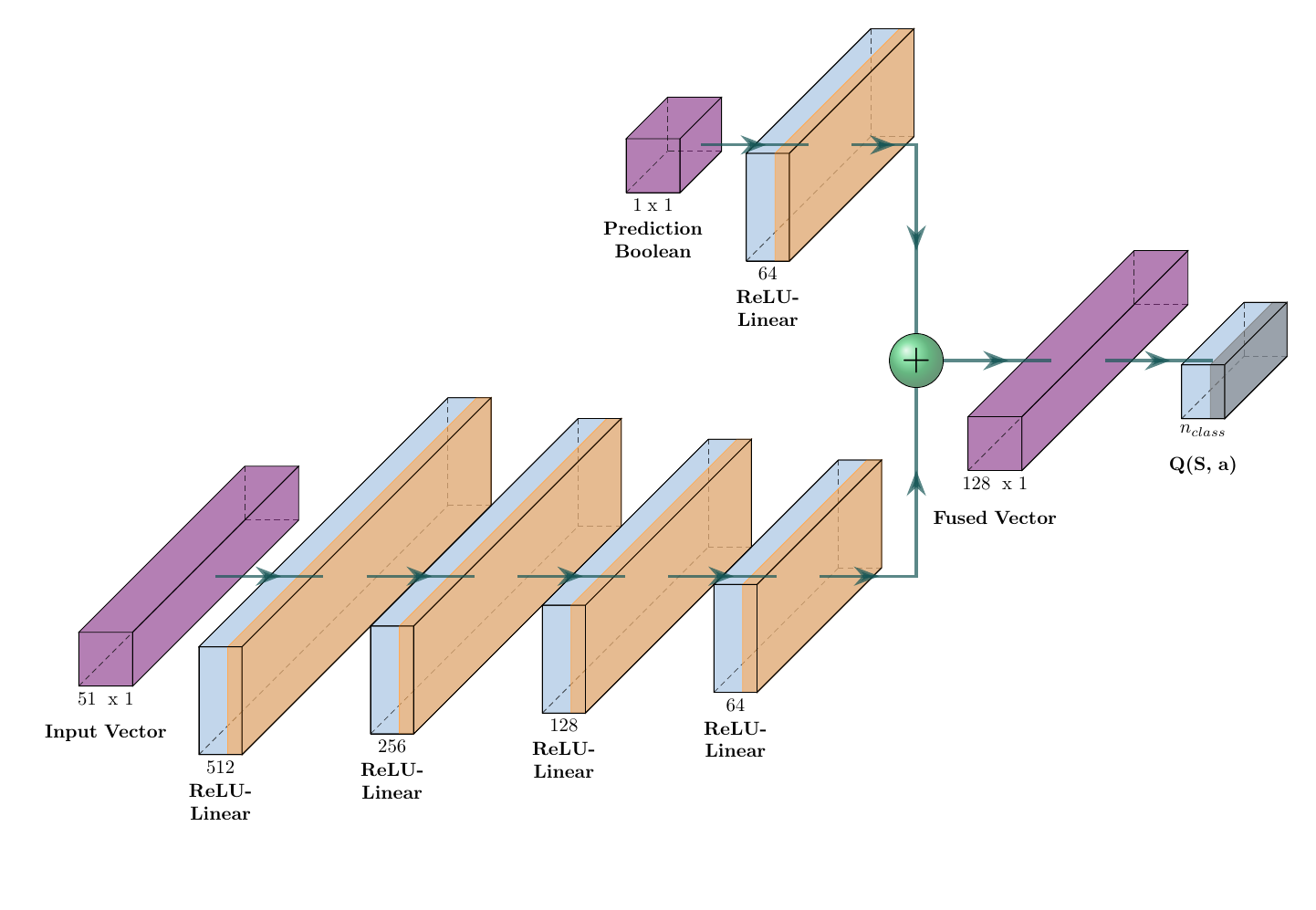}
    \caption{The Multi-Input neural network architecture used for the Deep Reinforcement Learning setting. The two components of the state are initially processed in separate branches to ensure compatibility.  }
    \label{3_4}
\end{figure}

\noindent
The DRL setting uses a neural network (functional modeling) to predict the Q-values of possible actions at a given state. The policy of action selection during each training step is based on the $\epsilon$-greedy strategy; each transition data ($s_t$, $a_t$, $R_{t+1}$, $s_{t+1}$) is stored in a memory buffer. The network optimization step proceeds by sampling a batch of transitions from the memory buffer and estimating the target Q-value using equation \ref{eq6}. Now a simple Mean Squared Error objective can be used to compare the error in the Q-value estimate as modeled by the network. Note that the hyperparameter setting for the optimization setting is the same as that used in the other classification settings, which enables formal comparison between all settings used. Illustrated in Figure \ref{3_4}, the network architecture is slightly varied to accommodate for the boolean state variable.

\section{Experiments and Results}

\subsection{Data Collection}
A set of molecular parameters were chosen for each of the peroxy radicals of particular VOC, which reacts with NO, substrate from the minimum energy geometries of the peroxy radical optimized with G3B3 quantum composite method by using Gaussian16 program suite \cite{frisch2016gaussian16}. A total of 51 global descriptors were chosen in this work as input parameters to represent an overall structural representation of the peroxy radical. These molecular parameters included the number of carbon(nC), the number of hydrogen(nH), and various other atoms present in the molecular structure of peroxy radical. The parameters also included molecular distance edge(MDEC), fraction of rotatable bonds(RotBtFrac), partial positive surface area (PPSA), and various other global descriptors obtained using PaDEL software \cite{yap2011padel}. The parameters were chosen to make the model efficient in predicting the reaction outcome with significant accuracy.

\subsection{Data Preparation}

The dataset from experimental analysis contained 91 data points. We assumed that the data distribution obtained was noise-free and that possible experimental biases were at a minimum. Considering the value range of the dependent variable ($Y$), we re-scaled using standard log scaling: 

\begin{equation}
    Y = \log{\left(\frac{Y-Y_{min}}{Y_{max}-Y_{min}}\right)}
    \label{log_sc}
\end{equation}

\noindent
The train-test split for regression was 81:10, where the test data was sampled at random from the dataset. For the purposes of binary classification, the class distribution obtained after the clustering procedure was 41:50 (high yield: low yield). The train-test split was chosen to be 71:20. For testing, 10 data-points were chosen at random from each class.

\subsection{Model Comparison}

\begin{figure}[!ht]
\hspace*{-1.5cm}
\begin{minipage}{.6\linewidth}
\centering
\subfloat[Vanilla Classification]{\label{main:a}\includegraphics[width=\textwidth, height=6.5cm]{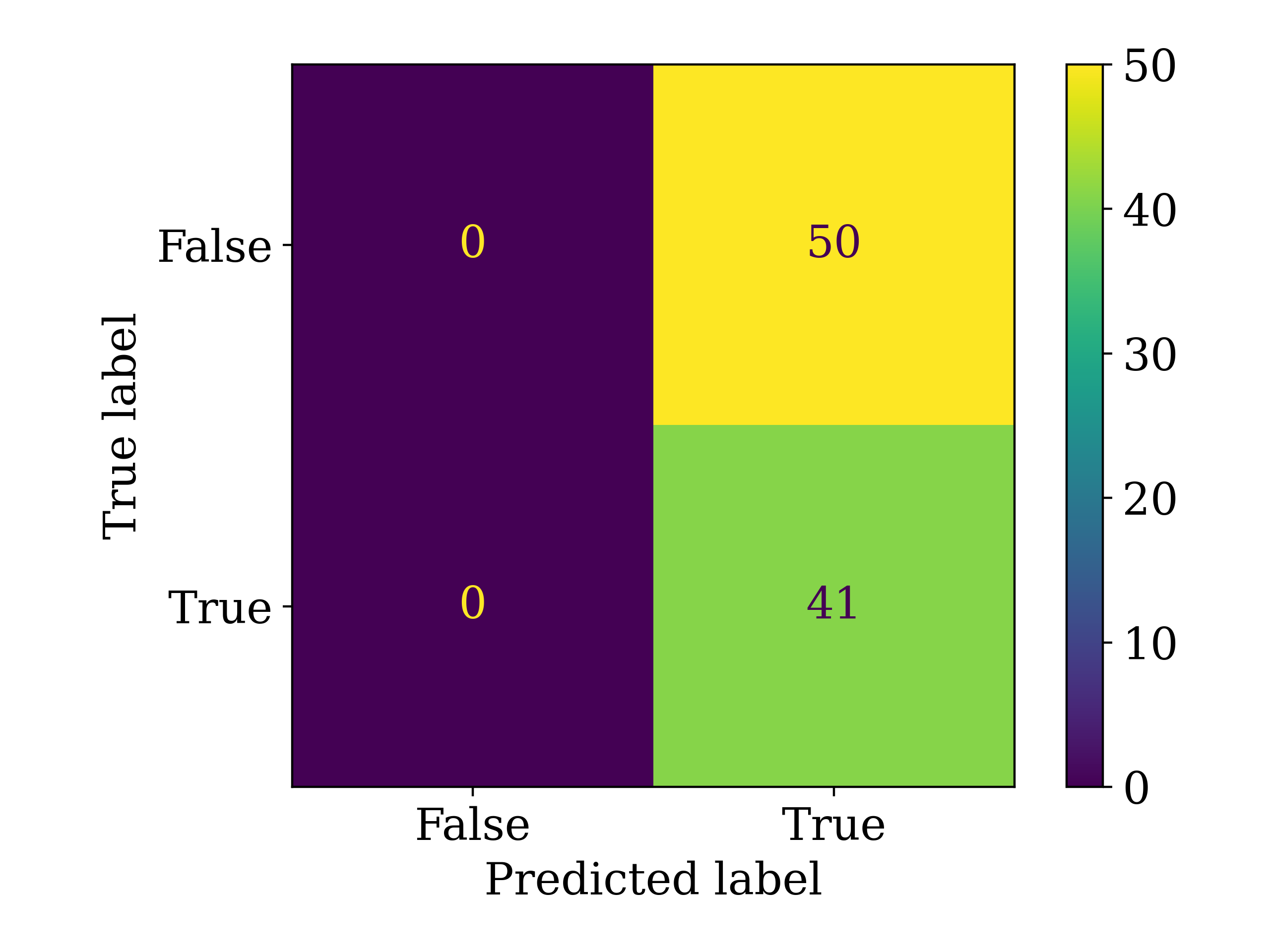}}
\end{minipage}%
\begin{minipage}{.6\linewidth}
\centering
\subfloat[Representational Learning]{\label{main:b}\includegraphics[width=\textwidth, height=6.5cm]{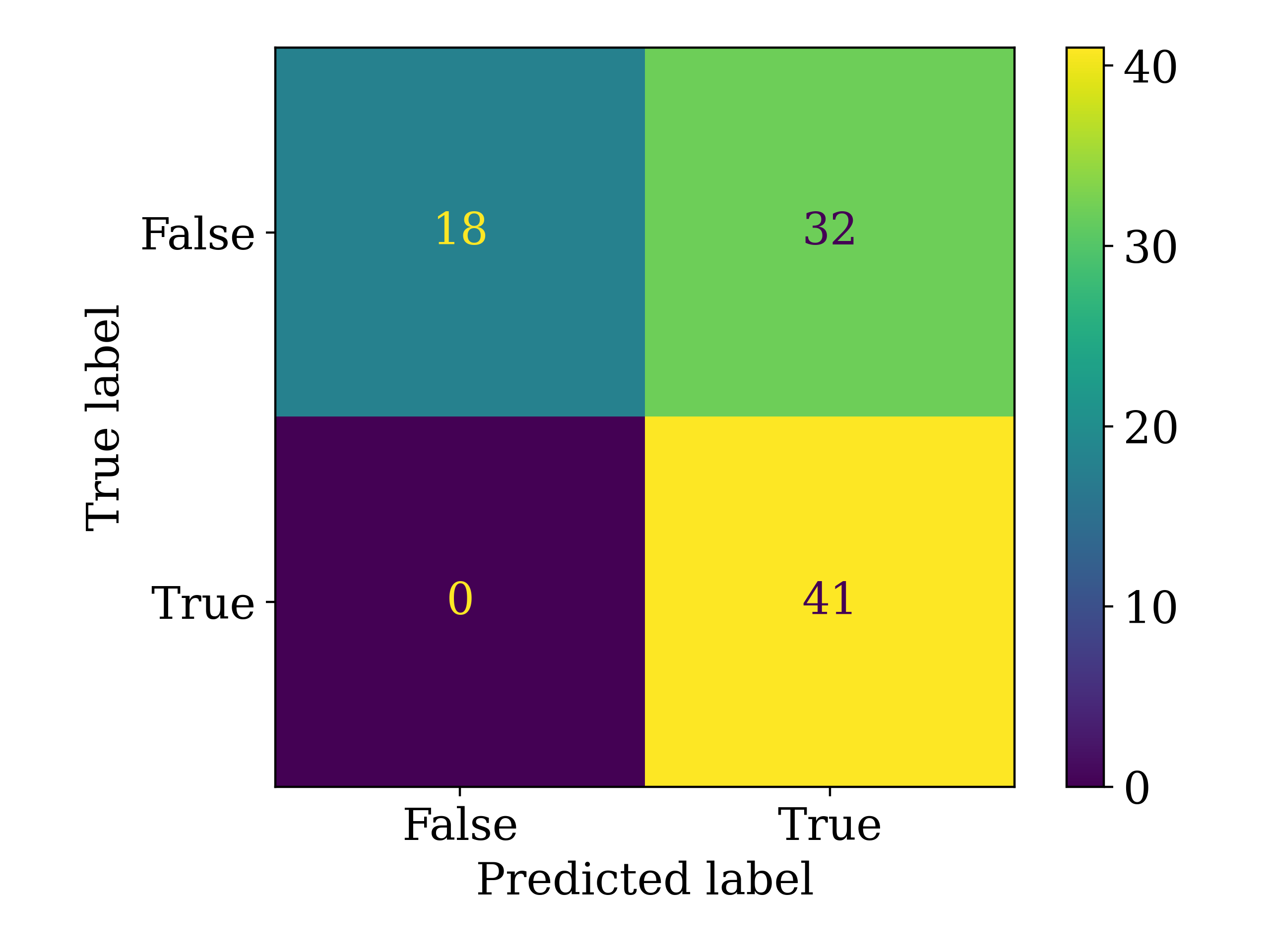}}
\end{minipage}\par\medskip
\centering
\subfloat[RL based Classification]{\label{main:c}\includegraphics[width=0.6\textwidth, height=6.5cm]{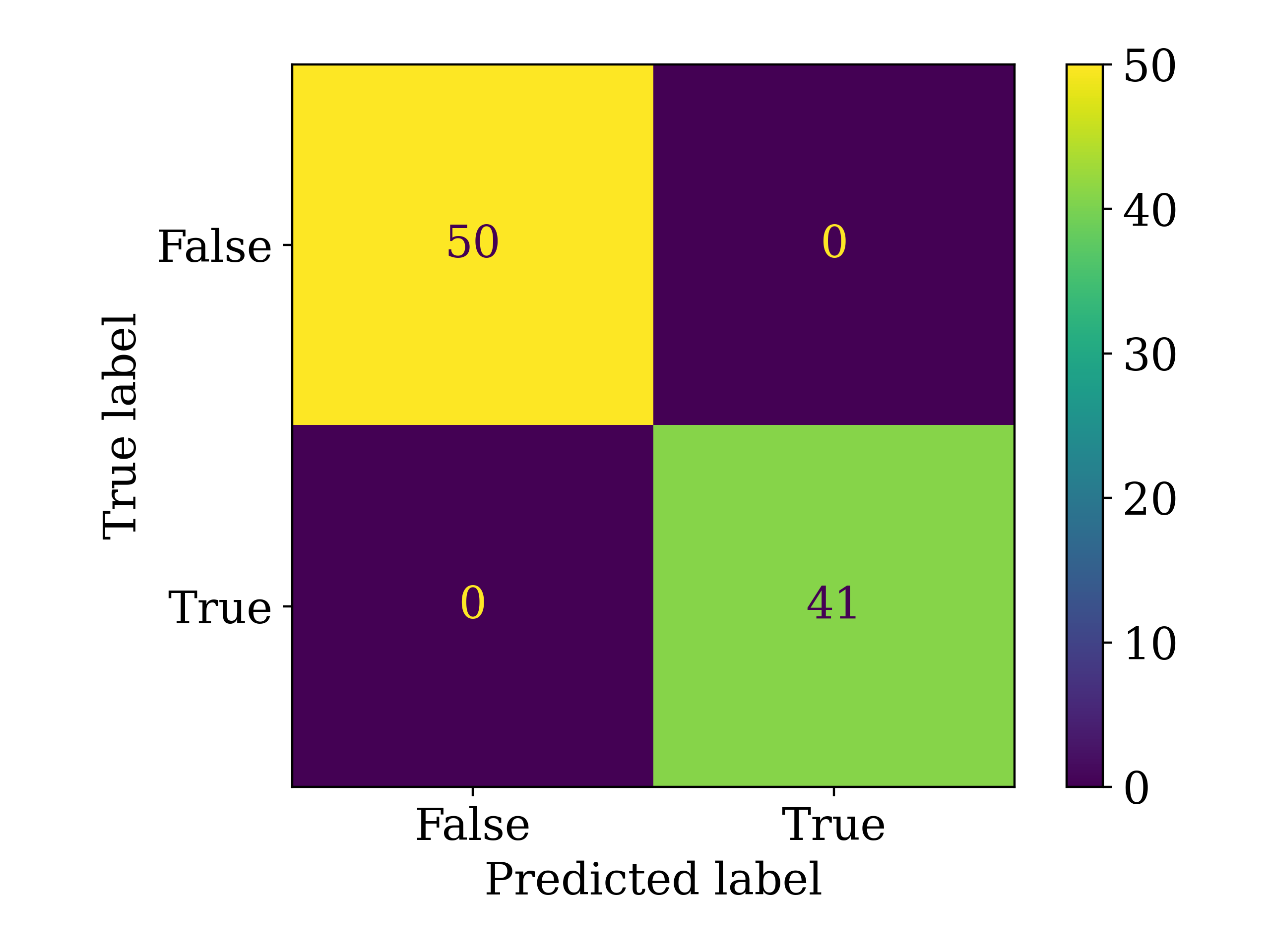}}
\caption{Average confusion matrices (ceiling integer value) obtained for corresponding models on full dataset evaluation, post model training in each proposed setting.}
\label{fig:main}
\end{figure}

\noindent
We began our analysis by cross-comparing the classification metrics obtained by the three different settings. The confusion matrices obtained after runs on a fixed hyperparamter (tuned) setting are shown in Figure \ref{fig:main}. Note that the RL-based setting was able to converge nominally in the dataset, whereas the Vanilla classification setting showed under-fitting behavior. Our study aimed to investigate the significance of the intricate analysis perspectives brought in through neural networks. The variable importance of the best-performing model is further analyzed using the Integrated Gradients (IGs) method to obtain critical insights into the analysis perspective brought in through predictive modeling of the data. The variable importance obtained from IGs is plotted out for each variable in possible ranges (limited data samples) as shown in Appendix \ref{app:1}. 

\begin{figure}[h!]
    \hspace*{-1.5cm}
    \centering
    \includegraphics[width=17cm, height=6cm]{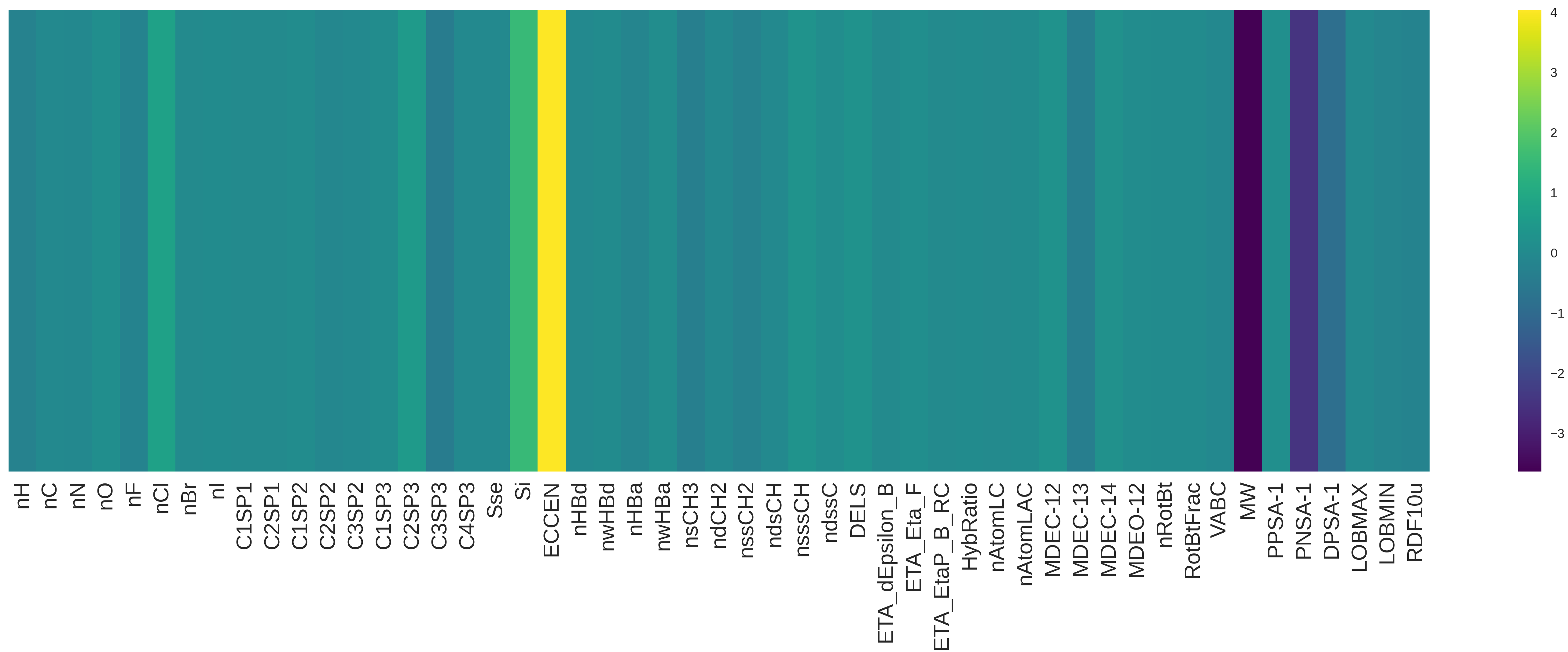}
    \caption{Integrated gradient map outlining the variable importance obtained.}
    \label{fig:ig}
\end{figure}

\noindent
We cross-compare the trends obtained from our method with those of existing literature to validate and obtain new insights that may open doors to possible research directions in the field of chemistry.

\section{Discussion}

\begin{figure}
 \begin{subfigure}{0.49\textwidth}
     \includegraphics[width=\textwidth]{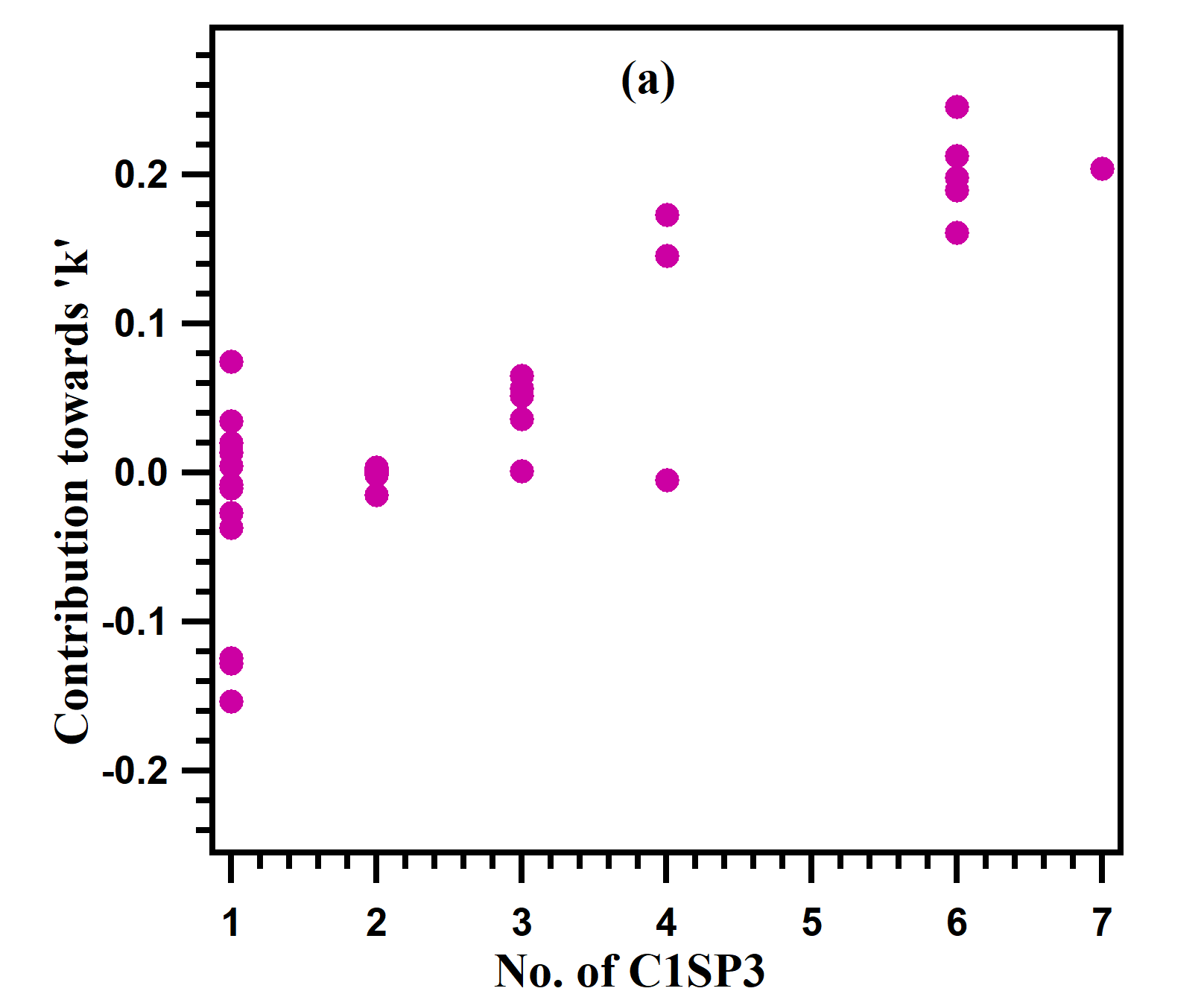}
     \label{fig:ig_comp:a}
 \end{subfigure}
 \hfill
 \begin{subfigure}{0.49\textwidth}
     \includegraphics[width=\textwidth]{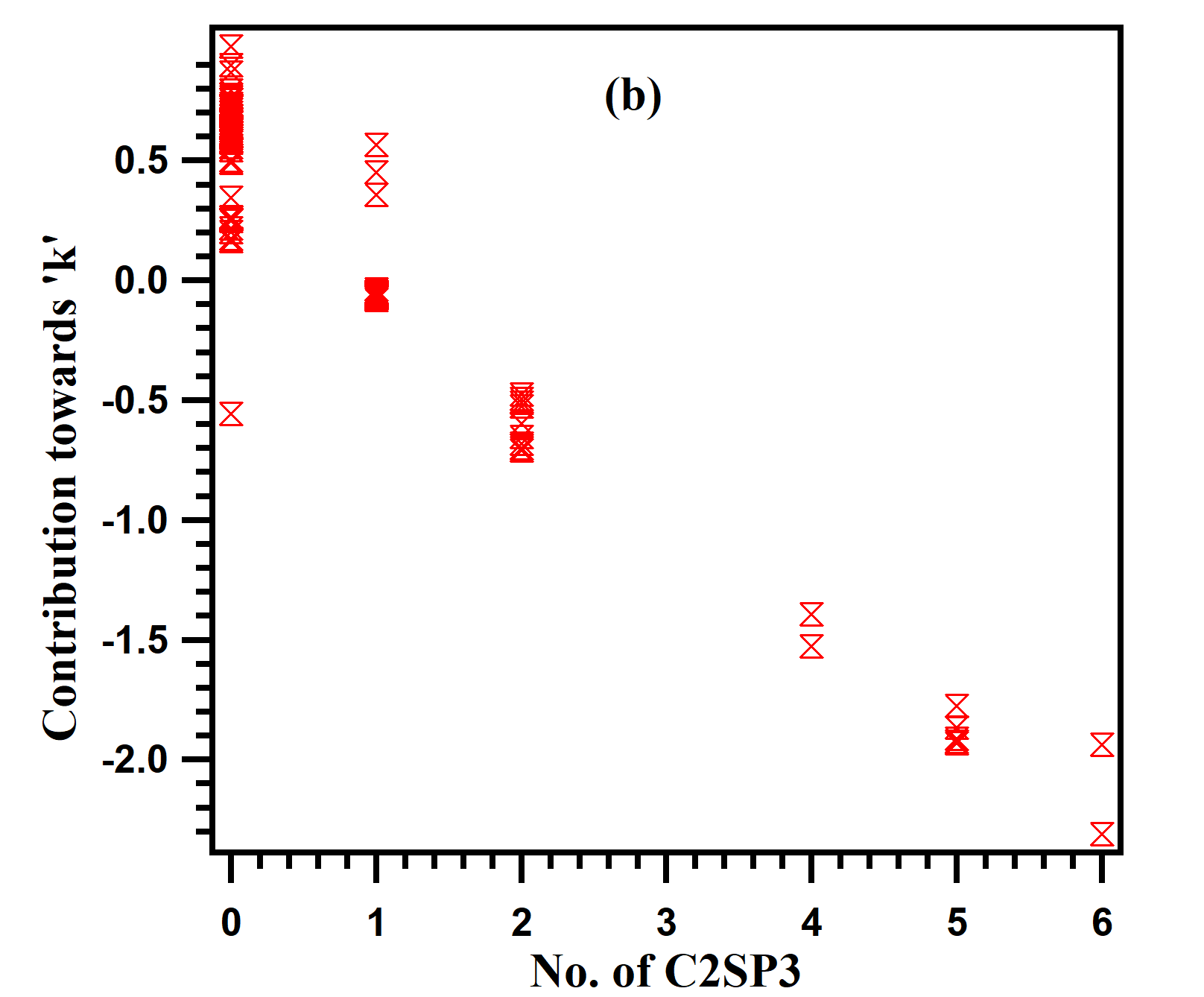}
     \label{fig:ig_comp:b}
 \end{subfigure}
 
 \begin{subfigure}{0.49\textwidth}
     \includegraphics[width=\textwidth]{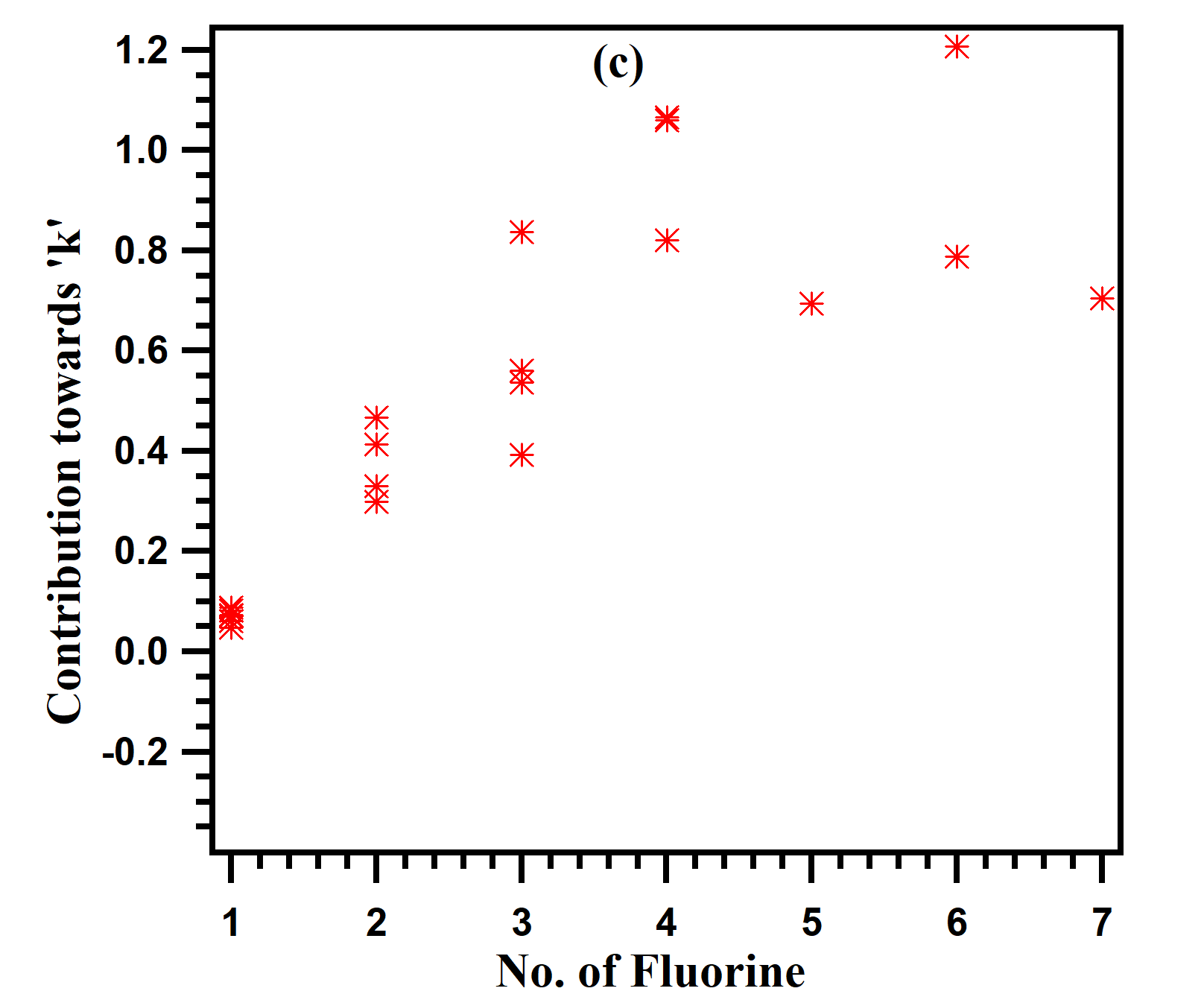}
     \label{fig:ig_comp:c}
 \end{subfigure}
 \hfill
 \begin{subfigure}{0.49\textwidth}
     \includegraphics[width=\textwidth]{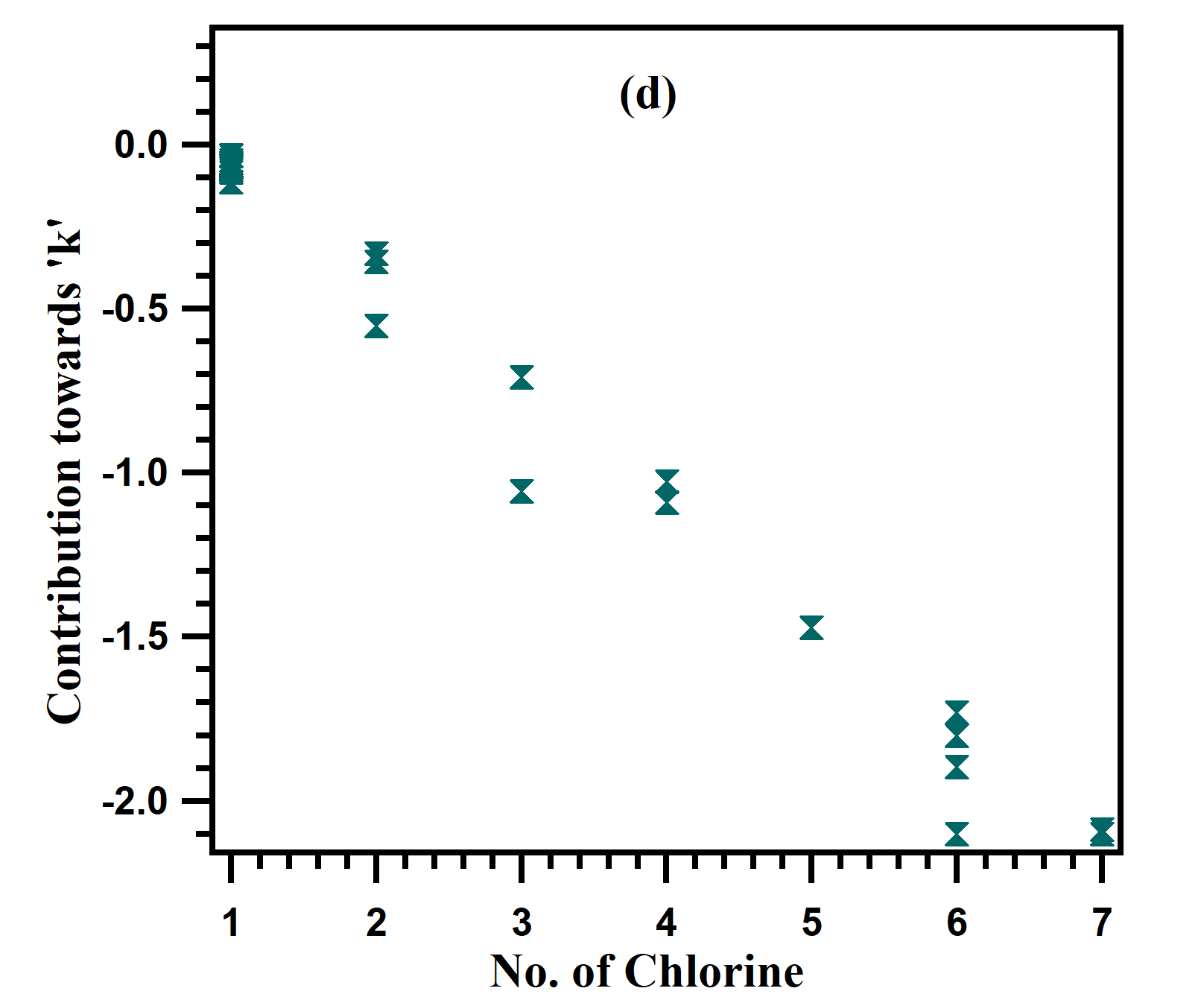}
     \label{fig:ig_comp:d}
 \end{subfigure}

 \medskip
 \begin{subfigure}{0.49\textwidth}
     \includegraphics[width=\textwidth]{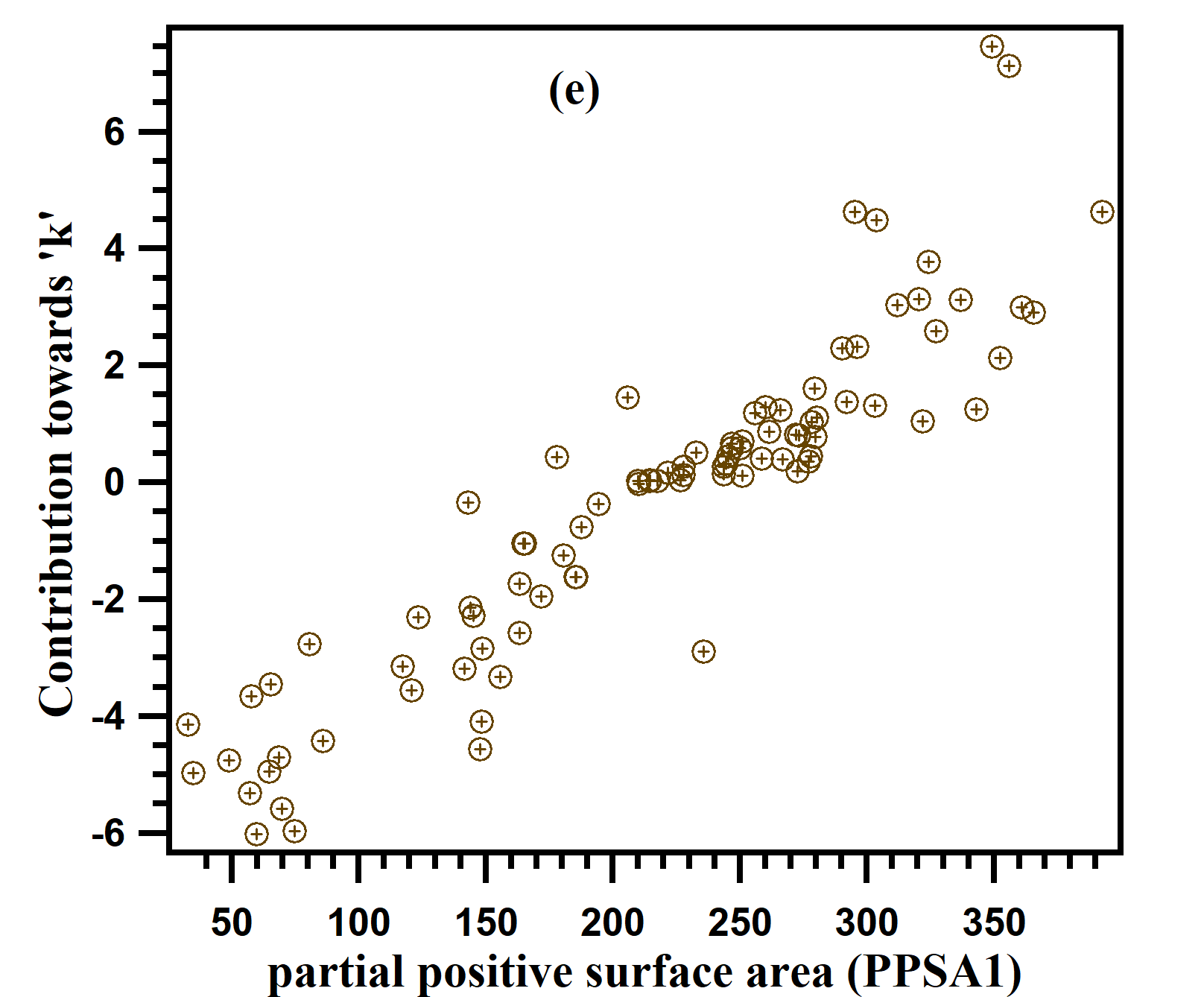}
     \label{fig:ig_comp:e}
 \end{subfigure}
 \hfill
 \begin{subfigure}{0.49\textwidth}
     \includegraphics[width=\textwidth]{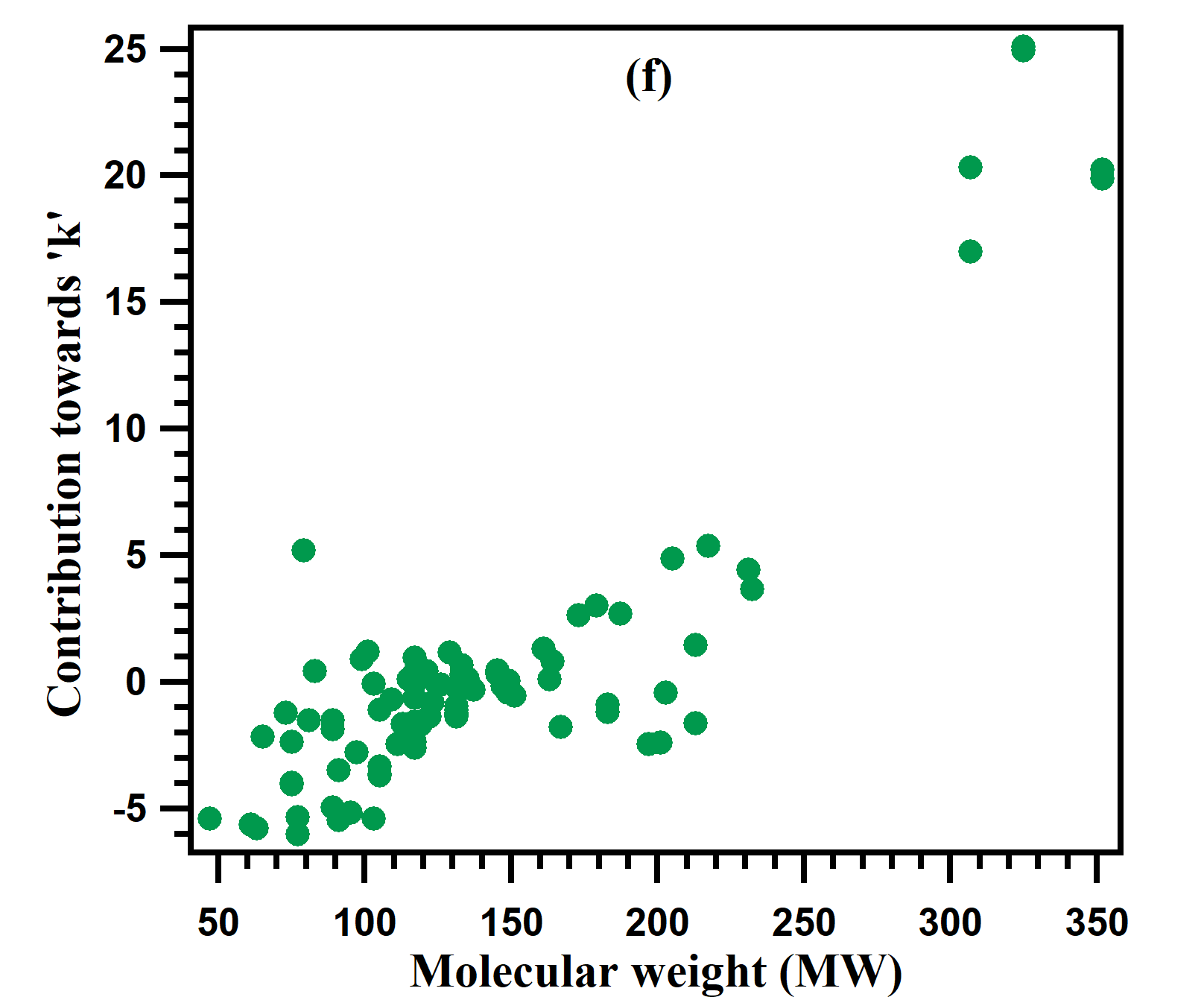}
     \label{fig:ig_comp:f}
 \end{subfigure}

 \caption{The contribution trends of various important chemical descriptors towards the rate constant "k" value.}
 \label{fig:ig_comp}

\end{figure}

We discovered several correlations between the input molecular descriptors and the target variable i.e., rate constant \textit{k}. Different geometric descriptors were used to describe the structural feature of different peroxy radicals and their interaction with NO. It is observed that the contribution toward \textit{k} is approximately linearly positively or negatively correlated with the number of carbon atoms in the peroxy radical and the type of bonding they were involved in, as shown in Figures \ref{fig:ig_comp}(a) and \ref{fig:ig_comp}(b). We observed that the bonding pattern and hybridization of the carbon atom also play an important role in determining \textit{k} for the studied class of reaction.

\noindent

In Figure \ref{fig:ig_comp}(a), the descriptor C1SP3 follows a positive correlation with rate constant \textit{k} which indicates that the contribution towards \textit{k} value increases with increasing number of SP\textsuperscript{3}\ "C" present in the molecule which are attached to only one other "C" atom. The increased number of terminal -CH\textsubscript{3}\ group increases the electron density on the peroxy radical as a result of +I inductive effect, which contributes towards the reactivity. Figure \ref{fig:ig_comp}(b) indicates a negative correlation between the descriptor C2SP3 and its contribution towards the "k" value. The increase in SP\textsuperscript{3}\ "C" attached to two other "C" i.e., -CH\textsubscript{2}\ group leads to a decrease in the contribution towards the "k" value. Here the +I effect of the terminal -CH\textsubscript{3}\ group decreases with an increase in the value of C2SP3 which causes the contribution to the "k" value to decline. 

The number of halogens also appears to correlate linearly \textit{k}, as seen in Figure \ref{fig:ig_comp}(c and d). In Figure \ref{fig:ig_comp}(c), we see that the contribution towards the "k" value increases with an increase in the number of Fluorine atoms present in the peroxy radical. On the other hand, in Figure \ref{fig:ig_comp}(d), the contribution towards "k" value decreases with increasing number of Cl atoms present in the peroxy radicals. Presumably, halogen bonding between the Cl atom and NO diminishes the interaction of NO with the peroxy radical and the feasibility of the reaction. In halogen bonding, a positive region, called a $\sigma$-hole, forms around the surface of "Cl." The $\sigma$-hole is responsible for the existence of halogen bonding of the "Cl" atom, whereas the halogen bonding is not shown by the "F" atom due to its higher electronegativity and limited polarizability \cite{esrafili2012investigation}. "F" exhibits a very strong -I effect, which contributes to the higher stability of the reactive complex produced between the peroxy radical and NO. We infer that the enhanced stability underlies the positive correlation between number of "F" and the "k" value. 

We also examined PPSA1, which is the partial positive surface area, defined by: 
\begin{equation} 
PPSA1 =  \Sigma(SA_i) \label{eq8}
\end{equation}
where ($\text{SA}_i$) is the surface contribution of $\text{i}_{th}$ positive atom in the peroxy radical. Figure \ref{fig:ig_comp}(e) illustrates the effect of PPSA1 by showing how the contribution to "k" increases as the peroxy radicals' PPSA1 value rises. 

Finally, it is evident in Figure \ref{fig:ig_comp}(f) that as the peroxy radical's molecular weight (MW) rises, so does its contribution to the "k" value.

\section{Conclusion and Future Work}

We have shown that deep learning with a basic artificial neural network can reliably identify key contributing factors to free radical reaction kinetics in the troposphere. In particular, deep reinforcement learning is a data-efficient technique that permits accurate algorithms when only a small amount of training data is present. In the present case, as is often the case, only a small amount -- by deep learning standards -- of experimental data was available for analysis. Nevertheless, as we have seen notably in the realm of medical images, deep reinforcement learning can learn in a generalized fashion even in the presence of limited training data. The deep reinforcement learning-based method used here is able to predict the range of kinetics and is able to generalize the predictive ability of the neural network to predictive kinetics of peroxy radicals with a testing set  accuracy of 100\%. In this study, we report a few observed correlations between the descriptors and their influence on the rate constant "k" value. In particular, the correlations obtained for the molecular descriptors such as number/type of halogens present in the peroxy radical as well as physical descriptors like PPSA1 and Molecular weight (MW) may provide a fresh perspective for atmospheric research on the kinetic behavior of various radicals. Future work will extend the current approach to use another small-data method: evolutionary strategies. Like deep reinforcement learning, evolutionary strategies have succeeded in image classification when only a few training images are present. We expect to obtain similar, intuitively consistent trends as we have in the presented work for deep reinforcement learning. We will in future work extend the approach to study related chemical systems in environmental science and systems biology. 

\section{Acknowledgment}

The authors SN and BR thank the High-Performance Computing Environment (HPCE) of IITM for providing the AQUA supercluster, which performed the calculations. SN thanks IITM for the fellowship. The authors HS and JS gratefully acknowledge grant support from the Radiological Society of North America, the American Society of Neuroradiology, as well as internal grant support from Memorial Sloan Kettering Cancer Center.

\section{Conflicts of Interests}

Two of the authors HS and JS are co-founders of Authera, Inc, a radiology artificial intelligence software company. Deep reinforcement learning is one of the approaches that our company uses, with associated patents received and pending.

\printbibliography

@article{yeo2000new,
  title={A new family of power transformations to improve normality or symmetry},
  author={Yeo, In-Kwon and Johnson, Richard A},
  journal={Biometrika},
  volume={87},
  number={4},
  pages={954--959},
  year={2000},
  publisher={Oxford University Press}
}

@inproceedings{zhang2018improved,
  title={Improved adam optimizer for deep neural networks},
  author={Zhang, Zijun},
  booktitle={2018 IEEE/ACM 26th International Symposium on Quality of Service (IWQoS)},
  pages={1--2},
  year={2018},
  organization={Ieee}
}

@inproceedings{schroff2015facenet,
  title={Facenet: A unified embedding for face recognition and clustering},
  author={Schroff, Florian and Kalenichenko, Dmitry and Philbin, James},
  booktitle={Proceedings of the IEEE conference on computer vision and pattern recognition},
  pages={815--823},
  year={2015}
}

@article{liashchynskyi2019grid,
  title={Grid search, random search, genetic algorithm: a big comparison for NAS},
  author={Liashchynskyi, Petro and Liashchynskyi, Pavlo},
  journal={arXiv preprint arXiv:1912.06059},
  year={2019}
}

@techreport{arthur2006k,
  title={k-means++: The advantages of careful seeding},
  author={Arthur, David and Vassilvitskii, Sergei},
  year={2006},
  institution={Stanford}
}

@article{khosla2020supervised,
  title={Supervised contrastive learning},
  author={Khosla, Prannay and Teterwak, Piotr and Wang, Chen and Sarna, Aaron and Tian, Yonglong and Isola, Phillip and Maschinot, Aaron and Liu, Ce and Krishnan, Dilip},
  journal={Advances in Neural Information Processing Systems},
  volume={33},
  pages={18661--18673},
  year={2020}
}

@article{stember2022deep,
  title={Deep reinforcement learning with automated label extraction from clinical reports accurately classifies 3D MRI brain volumes},
  author={Stember, Joseph Nathaniel and Shalu, Hrithwik},
  journal={Journal of Digital Imaging},
  pages={1--10},
  year={2022},
  publisher={Springer}
}

@article{mnih2015human,
  title={Human-level control through deep reinforcement learning},
  author={Mnih, Volodymyr and Kavukcuoglu, Koray and Silver, David and Rusu, Andrei A and Veness, Joel and Bellemare, Marc G and Graves, Alex and Riedmiller, Martin and Fidjeland, Andreas K and Ostrovski, Georg and others},
  journal={nature},
  volume={518},
  number={7540},
  pages={529--533},
  year={2015},
  publisher={Nature Publishing Group}
}

@article{sutton1999reinforcement,
  title={Reinforcement learning},
  author={Sutton, Richard S and Barto, Andrew G and others},
  journal={Journal of Cognitive Neuroscience},
  volume={11},
  number={1},
  pages={126--134},
  year={1999}
}

@article{watkins1992q,
  title={Q-learning},
  author={Watkins, Christopher JCH and Dayan, Peter},
  journal={Machine learning},
  volume={8},
  number={3},
  pages={279--292},
  year={1992},
  publisher={Springer}
}

@article{crutzen1979role,
  title={The role of NO and NO2 in the chemistry of the troposphere and stratosphere},
  author={Crutzen, Paul J},
  journal={Annual review of earth and planetary sciences},
  volume={7},
  pages={443--472},
  year={1979}
}

@article{yap2011padel,
  title={PaDEL-descriptor: An open source software to calculate molecular descriptors and fingerprints},
  author={Yap, Chun Wei},
  journal={Journal of computational chemistry},
  volume={32},
  number={7},
  pages={1466--1474},
  year={2011},
  publisher={Wiley Online Library}
}

@misc{frisch2016gaussian16,
  title={GAUSSIAN16. Revision A. 03. Gaussian Inc., Wallingford, CT, USA},
  author={Frisch, MJ and others},
  year={2016}
}

@article{atkinson2000atmospheric,
  title={Atmospheric chemistry of VOCs and NOx},
  author={Atkinson, Roger},
  journal={Atmospheric environment},
  volume={34},
  number={12-14},
  pages={2063--2101},
  year={2000},
  publisher={Elsevier}
}

@article{beker2019prediction,
  title={Prediction of Major Regio-, Site-, and Diastereoisomers in Diels--Alder Reactions by Using Machine-Learning: The Importance of Physically Meaningful Descriptors},
  author={Beker, Wiktor and Gajewska, Ewa P and Badowski, Tomasz and Grzybowski, Bartosz A},
  journal={Angewandte Chemie International Edition},
  volume={58},
  number={14},
  pages={4515--4519},
  year={2019},
  publisher={Wiley Online Library}
}

@article{singh2020unified,
  title={A unified machine-learning protocol for asymmetric catalysis as a proof of concept demonstration using asymmetric hydrogenation},
  author={Singh, Sukriti and Pareek, Monika and Changotra, Avtar and Banerjee, Sayan and Bhaskararao, Bangaru and Balamurugan, P and Sunoj, Raghavan B},
  journal={Proceedings of the National Academy of Sciences},
  volume={117},
  number={3},
  pages={1339--1345},
  year={2020},
  publisher={National Acad Sciences}
}

@article{jorner2021machine,
  title={Machine learning meets mechanistic modelling for accurate prediction of experimental activation energies},
  author={Jorner, Kjell and Brinck, Tore and Norrby, Per-Ola and Buttar, David},
  journal={Chemical Science},
  volume={12},
  number={3},
  pages={1163--1175},
  year={2021},
  publisher={Royal Society of Chemistry}
}

@article{lee2020graph,
  title={Graph theory-based reaction pathway searches and DFT calculations for the mechanism studies of free radical-initiated peptide sequencing mass spectrometry (FRIPS MS): a model gas-phase reaction of GGR tri-peptide},
  author={Lee, Jae-ung and Kim, Yeonjoon and Kim, Woo Youn and Oh, Han Bin},
  journal={Physical Chemistry Chemical Physics},
  volume={22},
  number={9},
  pages={5057--5069},
  year={2020},
  publisher={Royal Society of Chemistry}
}

@article{sanches2022evaluating,
  title={Evaluating and elucidating the reactivity of OH radicals with atmospheric organic pollutants: Reaction kinetics and mechanisms by machine learning},
  author={Sanches-Neto, Fl{\'a}vio O and Dias-Silva, Jefferson R and de Oliveira, Vitor M and Aquilanti, Vincenzo and Carvalho-Silva, Valter H},
  journal={Atmospheric Environment},
  volume={275},
  pages={119019},
  year={2022},
  publisher={Elsevier}
}

@article{ainsworth2012effects,
  title={The effects of tropospheric ozone on net primary productivity and implications for climate change},
  author={Ainsworth, Elizabeth A and Yendrek, Craig R and Sitch, Stephen and Collins, William J and Emberson, Lisa D},
  journal={Annual review of plant biology},
  volume={63},
  pages={637--661},
  year={2012},
  publisher={Annual Reviews}
}

@article{esrafili2012investigation,
  title={Investigation of H-bonding and halogen-bonding effects in dichloroacetic acid: DFT calculations of NQR parameters and QTAIM analysis},
  author={Esrafili, Mehdi D},
  journal={Journal of molecular modeling},
  volume={18},
  pages={5005--5016},
  year={2012},
  publisher={Springer}
}

@article{duce1983organic,
  title={Organic material in the global troposphere},
  author={Duce, RA and Mohnen, VA and Zimmerman, PR and Grosjean, D and Cautreels, W and Chatfield, R and Jaenicke, R and Ogren, JA and Pellizzari, ED and Wallace, GT},
  journal={Reviews of Geophysics},
  volume={21},
  number={4},
  pages={921--952},
  year={1983},
  publisher={Wiley Online Library}
}

@article{calvert1985chemical,
  title={Chemical mechanisms of acid generation in the troposphere},
  author={Calvert, Jack G and Lazrus, Allan and Kok, Gregory L and Heikes, Brian G and Walega, James G and Lind, John and Cantrell, Christopher A},
  journal={Nature},
  volume={317},
  number={6032},
  pages={27--35},
  year={1985},
  publisher={Nature Publishing Group UK London}
}

@article{highwood1998tropical,
  title={The tropical tropopause},
  author={Highwood, EJ and Hoskins, BJ},
  journal={Quarterly Journal of the Royal Meteorological Society},
  volume={124},
  number={549},
  pages={1579--1604},
  year={1998},
  publisher={Wiley Online Library}
}

@article{berntsen1997effects,
  title={Effects of anthropogenic emissions on tropospheric ozone and its radiative forcing},
  author={Berntsen, Terje Koren and Isaksen, Ivar SA and Myhre, G and Fuglestvedt, JS and Stordal, F and Larsen, T Alsvik and Freckleton, RS and Shine, Keith P},
  journal={Journal of Geophysical Research: Atmospheres},
  volume={102},
  number={D23},
  pages={28101--28126},
  year={1997},
  publisher={Wiley Online Library}
}

@article{zhang2003impacts,
  title={Impacts of anthropogenic and natural NOx sources over the US on tropospheric chemistry},
  author={Zhang, Renyi and Tie, Xuexi and Bond, Donald W},
  journal={Proceedings of the National Academy of Sciences},
  volume={100},
  number={4},
  pages={1505--1509},
  year={2003},
  publisher={National Acad Sciences}
}

@article{lamsal2011application,
  title={Application of satellite observations for timely updates to global anthropogenic NOx emission inventories},
  author={Lamsal, LN and Martin, RV and Padmanabhan, A and Van Donkelaar, A and Zhang, Q and Sioris, CE and Chance, K and Kurosu, TP and Newchurch, MJ},
  journal={Geophysical Research Letters},
  volume={38},
  number={5},
  year={2011},
  publisher={Wiley Online Library}
}

@article{wild2001indirect,
  title={Indirect long-term global radiative cooling from NOx emissions},
  author={Wild, Oliver and Prather, Michael J and Akimoto, Hajime},
  journal={Geophysical Research Letters},
  volume={28},
  number={9},
  pages={1719--1722},
  year={2001},
  publisher={Wiley Online Library}
}

@article{liu1987ozone,
  title={Ozone production in the rural troposphere and the implications for regional and global ozone distributions},
  author={Liu, SC and Trainer, M and Fehsenfeld, FC and Parrish, DD and Williams, EJ and Fahey, D Wo and H{\"u}bler, G and Murphy, P Co},
  journal={Journal of Geophysical Research: Atmospheres},
  volume={92},
  number={D4},
  pages={4191--4207},
  year={1987},
  publisher={Wiley Online Library}
}

@article{ganzeveld2002global,
  title={Global soil-biogenic NOx emissions and the role of canopy processes},
  author={Ganzeveld, LN and Lelieveld, J and Dentener, FJ and Krol, MC and Bouwman, AJ and Roelofs, G-J},
  journal={Journal of Geophysical Research: Atmospheres},
  volume={107},
  number={D16},
  pages={ACH--9},
  year={2002},
  publisher={Wiley Online Library}
}

@article{crutzen1988tropospheric,
  title={Tropospheric ozone: An overview},
  author={Crutzen, Paul J},
  journal={Tropospheric ozone: regional and global scale interactions},
  pages={3--32},
  year={1988},
  publisher={Springer}
}

@article{jaffe1967biological,
  title={The biological effects of ozone on man and animals},
  author={Jaffe, Louis S},
  journal={American Industrial Hygiene Association Journal},
  volume={28},
  number={3},
  pages={267--277},
  year={1967},
  publisher={Taylor \& Francis}
}

@article{iriti2008oxidative,
  title={Oxidative stress, the paradigm of ozone toxicity in plants and animals},
  author={Iriti, Marcello and Faoro, Franco},
  journal={Water, Air, and Soil Pollution},
  volume={187},
  pages={285--301},
  year={2008},
  publisher={Springer}
}

@article{witschi1999ozone,
  title={Ozone carcinogenesis revisited.},
  author={Witschi, Hanspeter and Espiritu, Imelda and Pinkerton, Kent E and Murphy, Kerry and Maronpot, Robert R},
  journal={Toxicological sciences: an official journal of the Society of Toxicology},
  volume={52},
  number={2},
  pages={162--167},
  year={1999}
}

@incollection{paige1999acute,
  title={Acute and chronic effects of ozone in animal models},
  author={Paige, Renee C and Plopper, Charles G},
  booktitle={Air pollution and health},
  pages={531--557},
  year={1999},
  publisher={Elsevier}
}

@article{fehsenfeld1992emissions,
  title={Emissions of volatile organic compounds from vegetation and the implications for atmospheric chemistry},
  author={Fehsenfeld, Fred and Calvert, Jack and Fall, Ray and Goldan, Paul and Guenther, Alex B and Hewitt, C Nicholas and Lamb, Brian and Liu, Shaw and Trainer, Michael and Westberg, Hal and others},
  journal={Global biogeochemical cycles},
  volume={6},
  number={4},
  pages={389--430},
  year={1992},
  publisher={Wiley Online Library}
}

@article{pinto2010plant,
  title={Plant volatile organic compounds (VOCs) in ozone (O 3) polluted atmospheres: the ecological effects},
  author={Pinto, Delia M and Blande, James D and Souza, Silvia R and Nerg, Anne-Marja and Holopainen, Jarmo K},
  journal={Journal of chemical ecology},
  volume={36},
  pages={22--34},
  year={2010},
  publisher={Springer}
}

@article{heiden2003emissions,
  title={Emissions of oxygenated volatile organic compounds from plants part I: emissions from lipoxygenase activity.},
  author={Heiden, AC and Kobel, K and Langebartels, C and Schuh-Thomas, G and Wildt, J},
  journal={Journal of atmospheric chemistry},
  volume={45},
  number={2},
  year={2003}
}

@article{wang2018attribution,
  title={Attribution of tropospheric ozone to NO x and VOC emissions: considering ozone formation in the transition regime},
  author={Wang, Peng and Chen, Yuan and Hu, Jianlin and Zhang, Hongliang and Ying, Qi},
  journal={Environmental science \& technology},
  volume={53},
  number={3},
  pages={1404--1412},
  year={2018},
  publisher={ACS Publications}
}

@article{oyama2000chemical,
  title={Chemical and catalytic properties of ozone},
  author={Oyama, S Ted},
  journal={Catalysis Reviews},
  volume={42},
  number={3},
  pages={279--322},
  year={2000},
  publisher={Taylor \& Francis}
}

@article{chen2022machine,
  title={Machine-learning-guided reaction kinetics prediction towards solvent identification for chemical absorption of carbonyl sulfide},
  author={Chen, Yuxiang and Liu, Chuanlei and Guo, Guanchu and Zhao, Yang and Qian, Cheng and Jiang, Hao and Shen, Benxian and Wu, Di and Cao, Fahai and Sun, Hui},
  journal={Chemical Engineering Journal},
  volume={444},
  pages={136662},
  year={2022},
  publisher={Elsevier}
}

@article{stocker2020machine,
  title={Machine learning in chemical reaction space},
  author={Stocker, Sina and Cs{\'a}nyi, G{\'a}bor and Reuter, Karsten and Margraf, Johannes T},
  journal={Nature communications},
  volume={11},
  number={1},
  pages={5505},
  year={2020},
  publisher={Nature Publishing Group UK London}
}

@article{meuwly2021machine,
  title={Machine learning for chemical reactions},
  author={Meuwly, Markus},
  journal={Chemical Reviews},
  volume={121},
  number={16},
  pages={10218--10239},
  year={2021},
  publisher={ACS Publications}
}

@article{mardt2018vampnets,
  title={VAMPnets for deep learning of molecular kinetics},
  author={Mardt, Andreas and Pasquali, Luca and Wu, Hao and No{\'e}, Frank},
  journal={Nature communications},
  volume={9},
  number={1},
  pages={5},
  year={2018},
  publisher={Nature Publishing Group UK London}
}

@article{lecun2015deep,
  title={Deep learning},
  author={LeCun, Yann and Bengio, Yoshua and Hinton, Geoffrey},
  journal={nature},
  volume={521},
  number={7553},
  pages={436--444},
  year={2015},
  publisher={Nature Publishing Group UK London}
}

@article{kollenz2020unravelling,
  title={Unravelling the kinetic model of photochemical reactions via deep learning},
  author={Kollenz, Philipp and Herten, Dirk-Peter and Buckup, Tiago},
  journal={The Journal of Physical Chemistry B},
  volume={124},
  number={29},
  pages={6358--6368},
  year={2020},
  publisher={ACS Publications}
}

@article{wu2022reaction,
  title={Reaction Kinetic Model Considering the Solvation Effect Based on the FMO Theory and Deep Learning},
  author={Wu, Xinyuan and Liu, Qilei and Zhao, Yujing and Zhang, Lei and Du, Jian},
  journal={Industrial \& Engineering Chemistry Research},
  volume={61},
  number={41},
  pages={15261--15272},
  year={2022},
  publisher={ACS Publications}
}

@article{zhong2020deep,
  title={A deep neural network combined with molecular fingerprints (DNN-MF) to develop predictive models for hydroxyl radical rate constants of water contaminants},
  author={Zhong, Shifa and Hu, Jiajie and Fan, Xudong and Yu, Xiong and Zhang, Huichun},
  journal={Journal of hazardous materials},
  volume={383},
  pages={121141},
  year={2020},
  publisher={Elsevier}
}

@article{zhong2021shedding,
  title={Shedding light on “Black Box” machine learning models for predicting the reactivity of HO radicals toward organic compounds},
  author={Zhong, Shifa and Zhang, Kai and Wang, Dong and Zhang, Huichun},
  journal={Chemical Engineering Journal},
  volume={405},
  pages={126627},
  year={2021},
  publisher={Elsevier}
}

\appendix

\section{Co-relations Obtained}
\label{app:1}

\begin{figure}[!ht]
\hspace*{-1.5cm}
\begin{minipage}{.6\linewidth}
\centering
\subfloat{\label{ap1:a}\includegraphics[width=\textwidth, height=6.5cm]{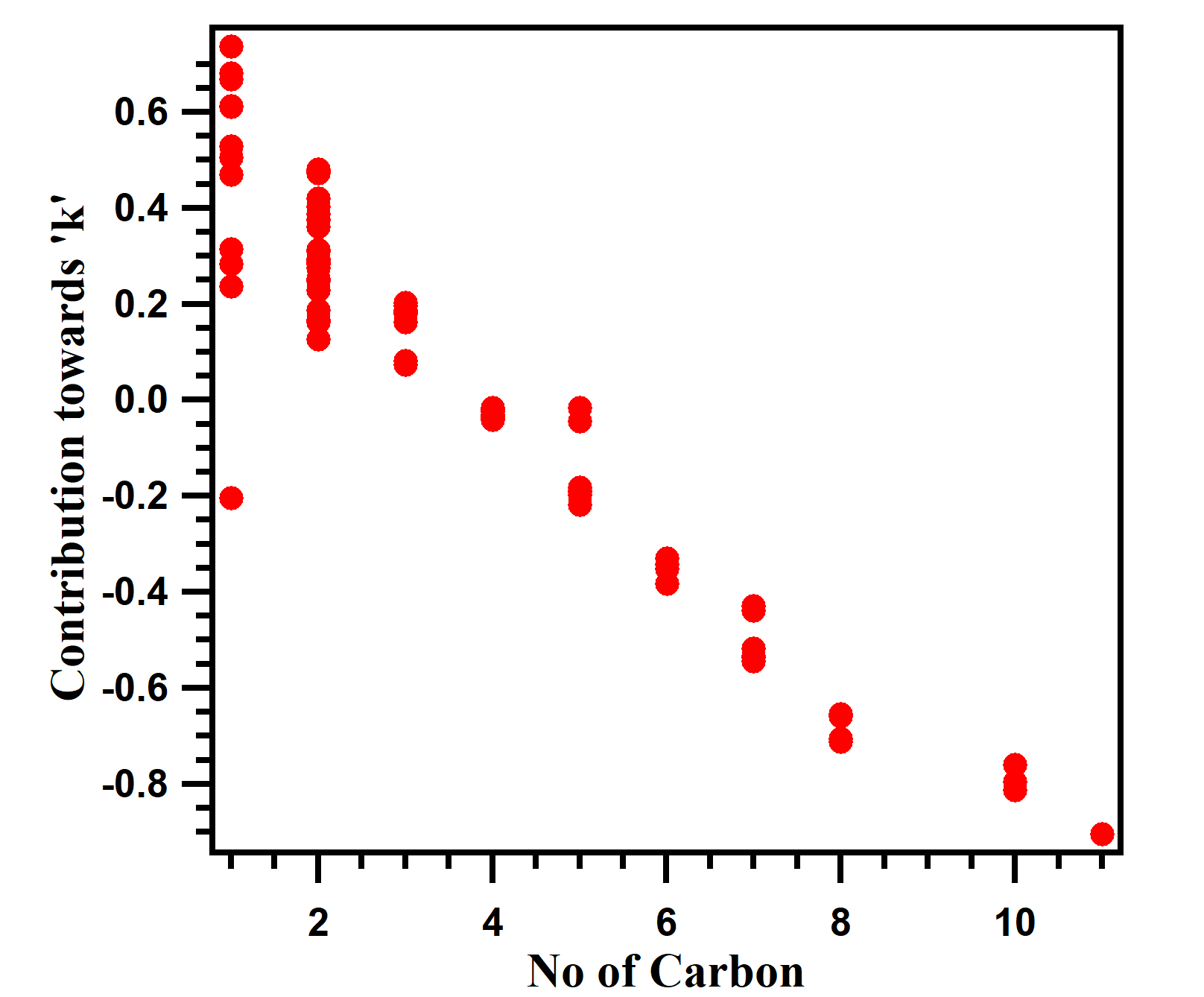}}
\end{minipage}%
\begin{minipage}{.6\linewidth}
\centering
\subfloat{\label{ap1:b}\includegraphics[width=\textwidth, height=6.5cm]{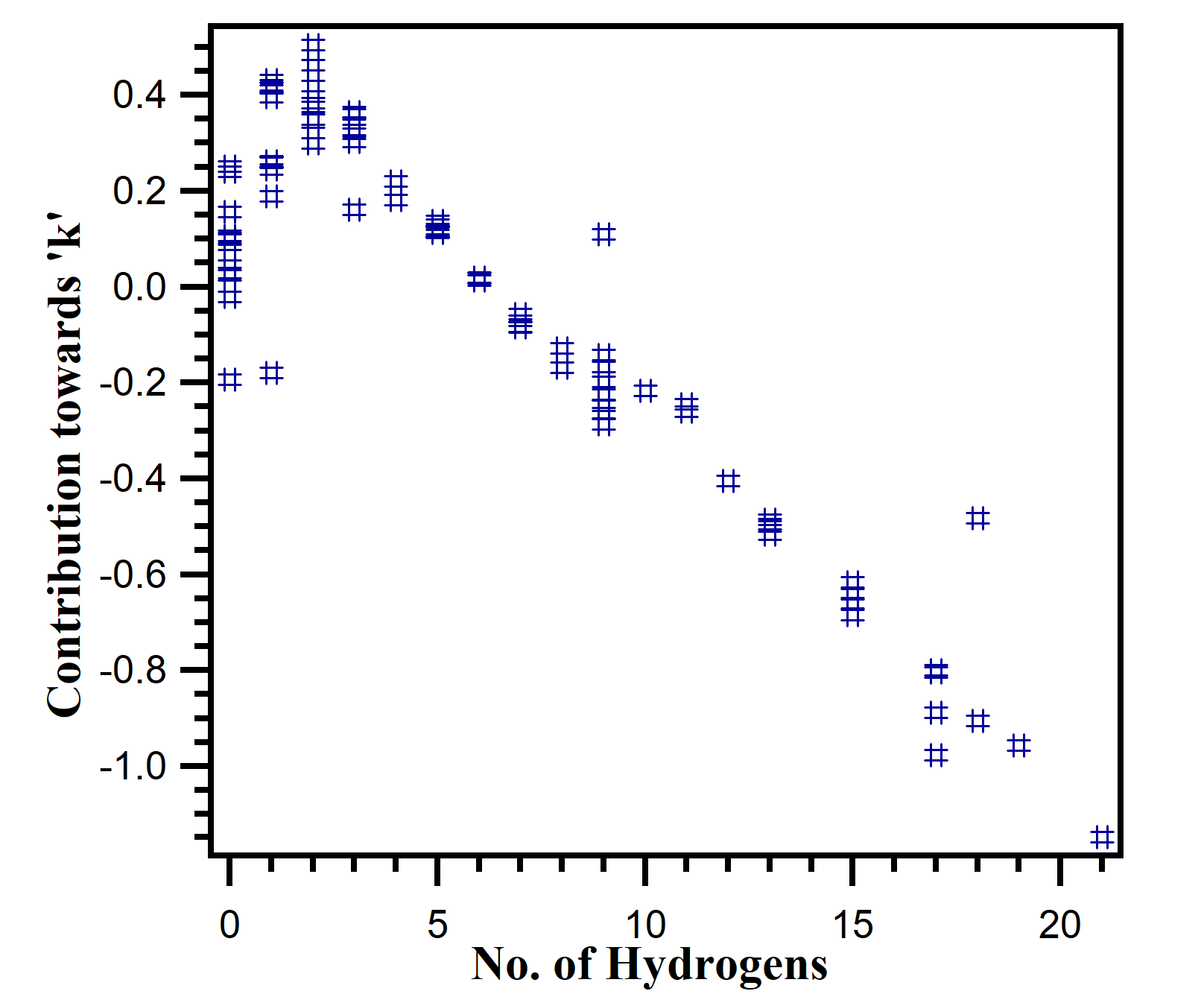}}
\end{minipage}\par\medskip
\label{fig:ap1}
\end{figure}

\begin{figure}[!ht]
\hspace*{-1.5cm}
\begin{minipage}{.6\linewidth}
\centering
\subfloat{\label{ap2:a}\includegraphics[width=\textwidth, height=6.5cm]{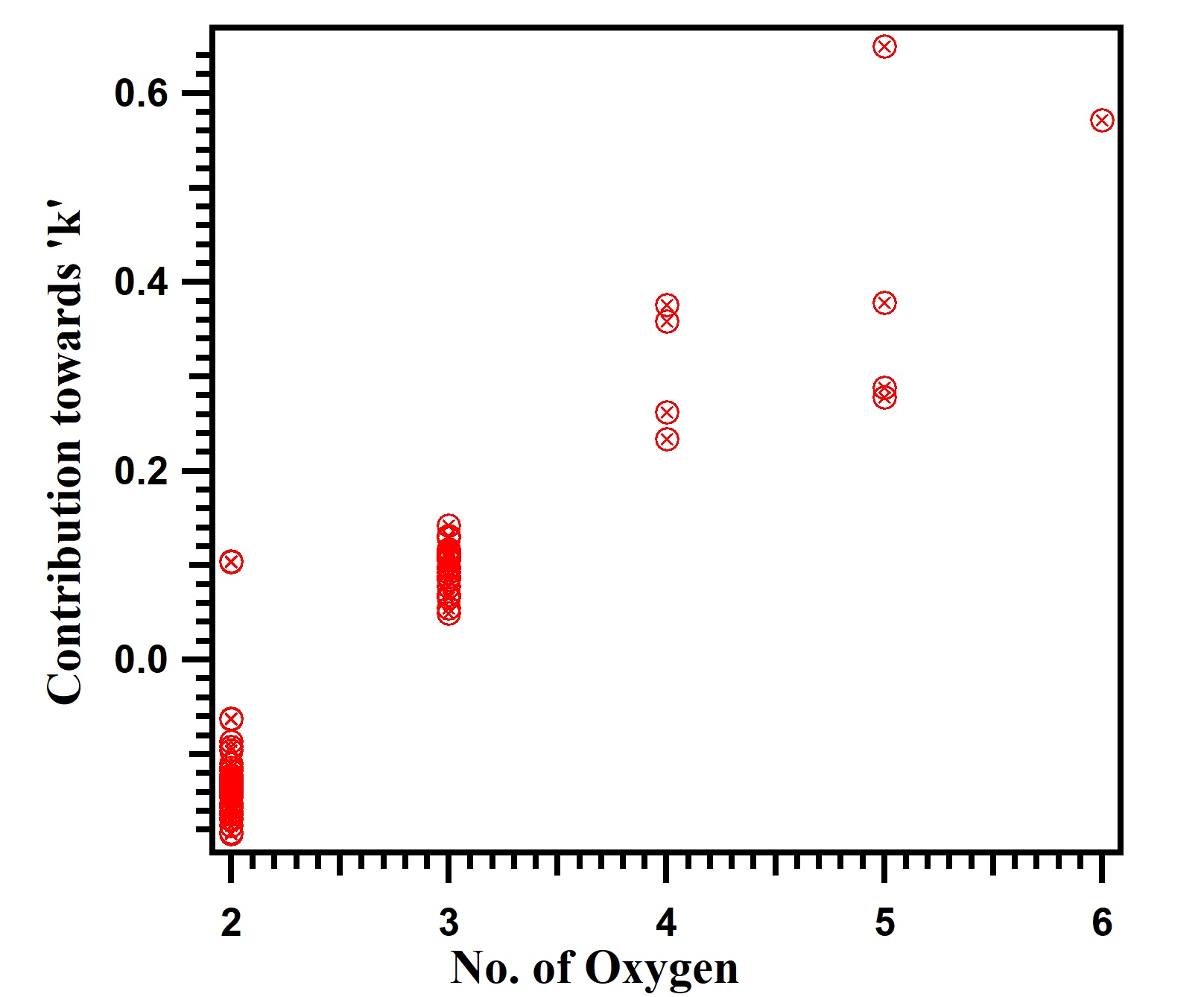}}
\end{minipage}%
\begin{minipage}{.6\linewidth}
\centering
\subfloat{\label{ap2:b}\includegraphics[width=\textwidth, height=6.5cm]{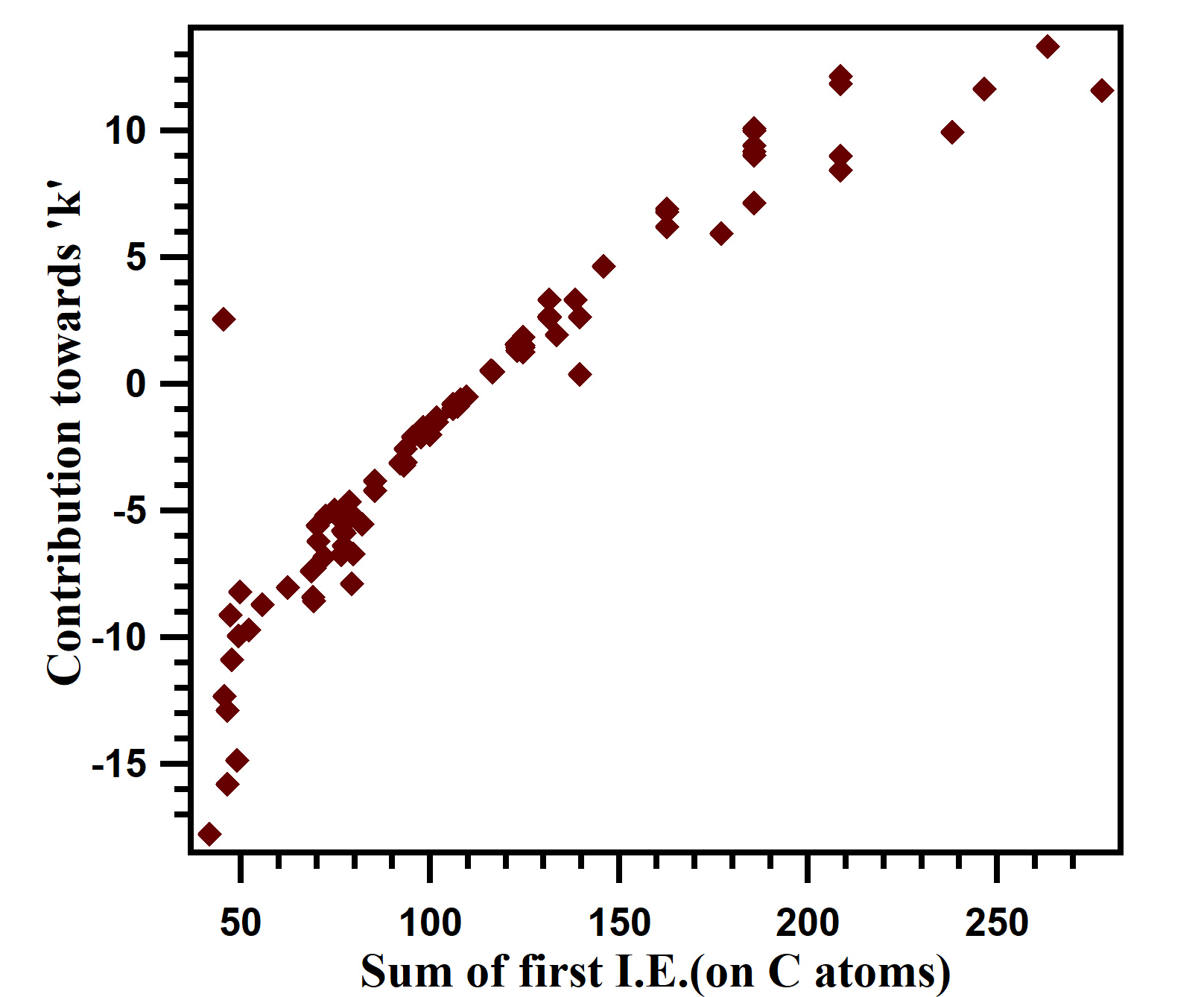}}
\end{minipage}\par\medskip
\label{fig:ap2}
\end{figure}

\begin{figure}[!ht]
\hspace*{-1.5cm}
\begin{minipage}{.6\linewidth}
\centering
\subfloat{\label{ap3:a}\includegraphics[width=\textwidth, height=6.5cm]{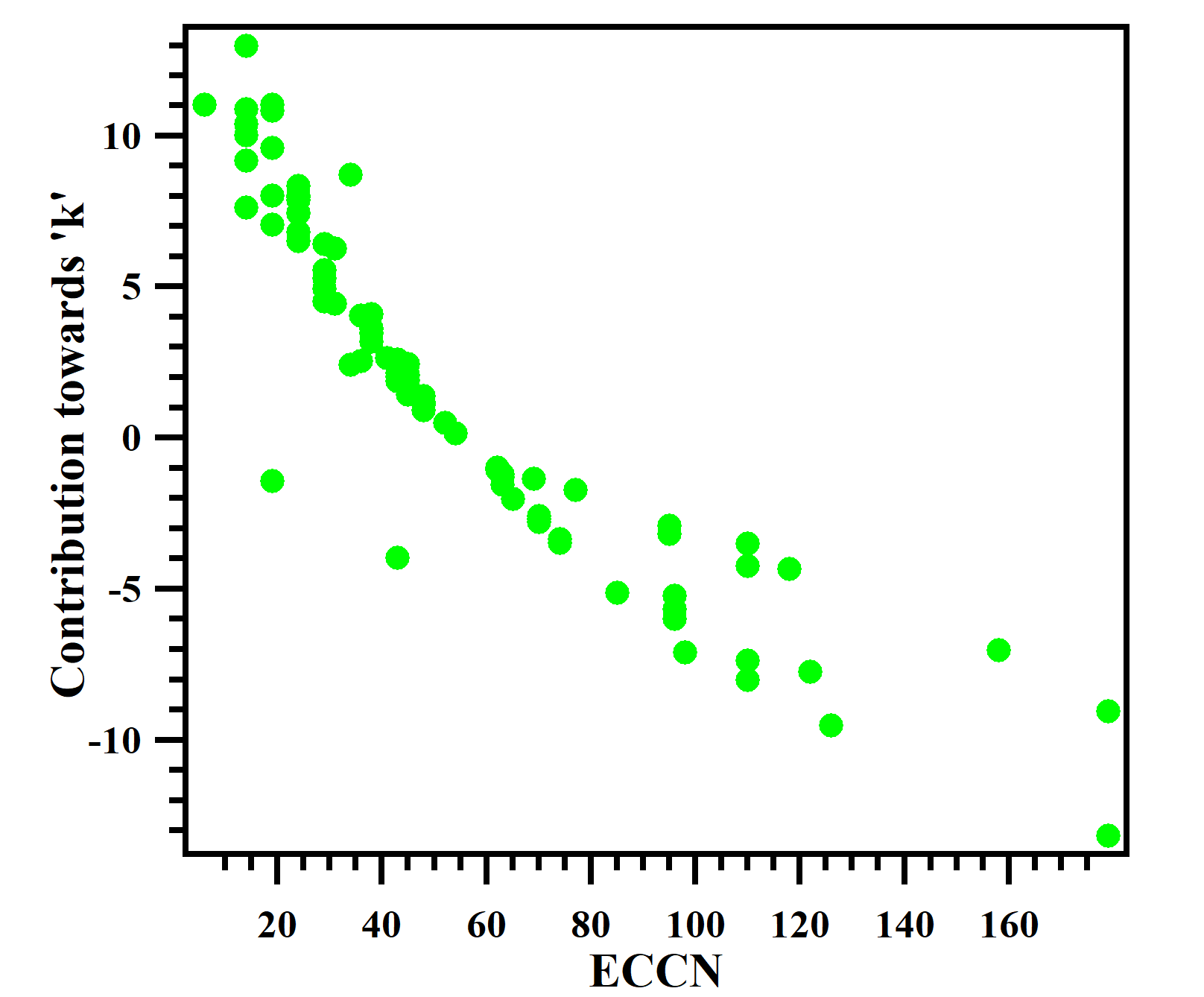}}
\end{minipage}%
\begin{minipage}{.6\linewidth}
\centering
\subfloat{\label{ap3:b}\includegraphics[width=\textwidth, height=6.5cm]{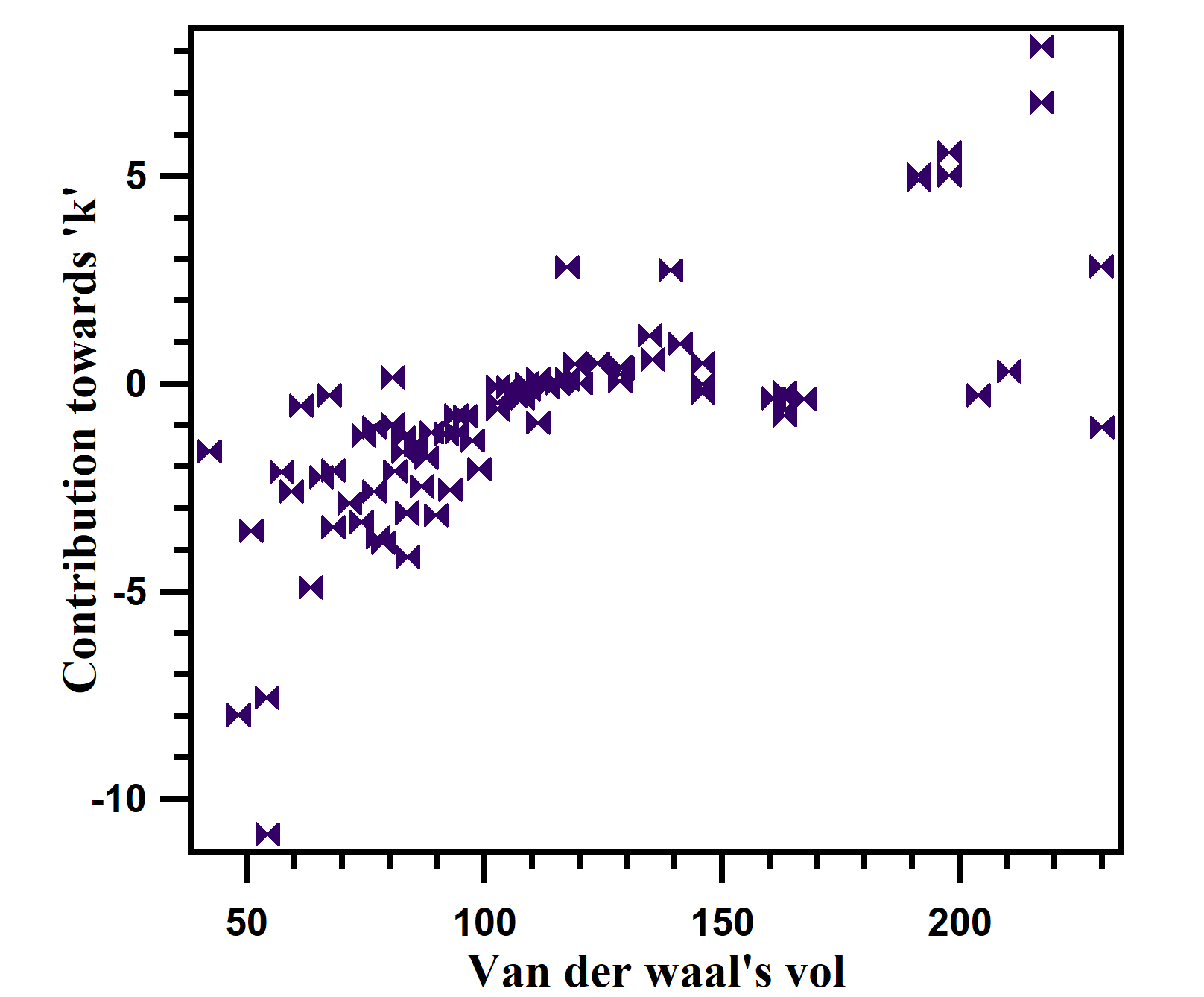}}
\end{minipage}\par\medskip
\label{fig:ap3}
\end{figure}

\begin{figure}[!ht]
\hspace*{-1.5cm}
\begin{minipage}{.6\linewidth}
\centering
\subfloat{\label{ap4:a}\includegraphics[width=\textwidth, height=6.5cm]{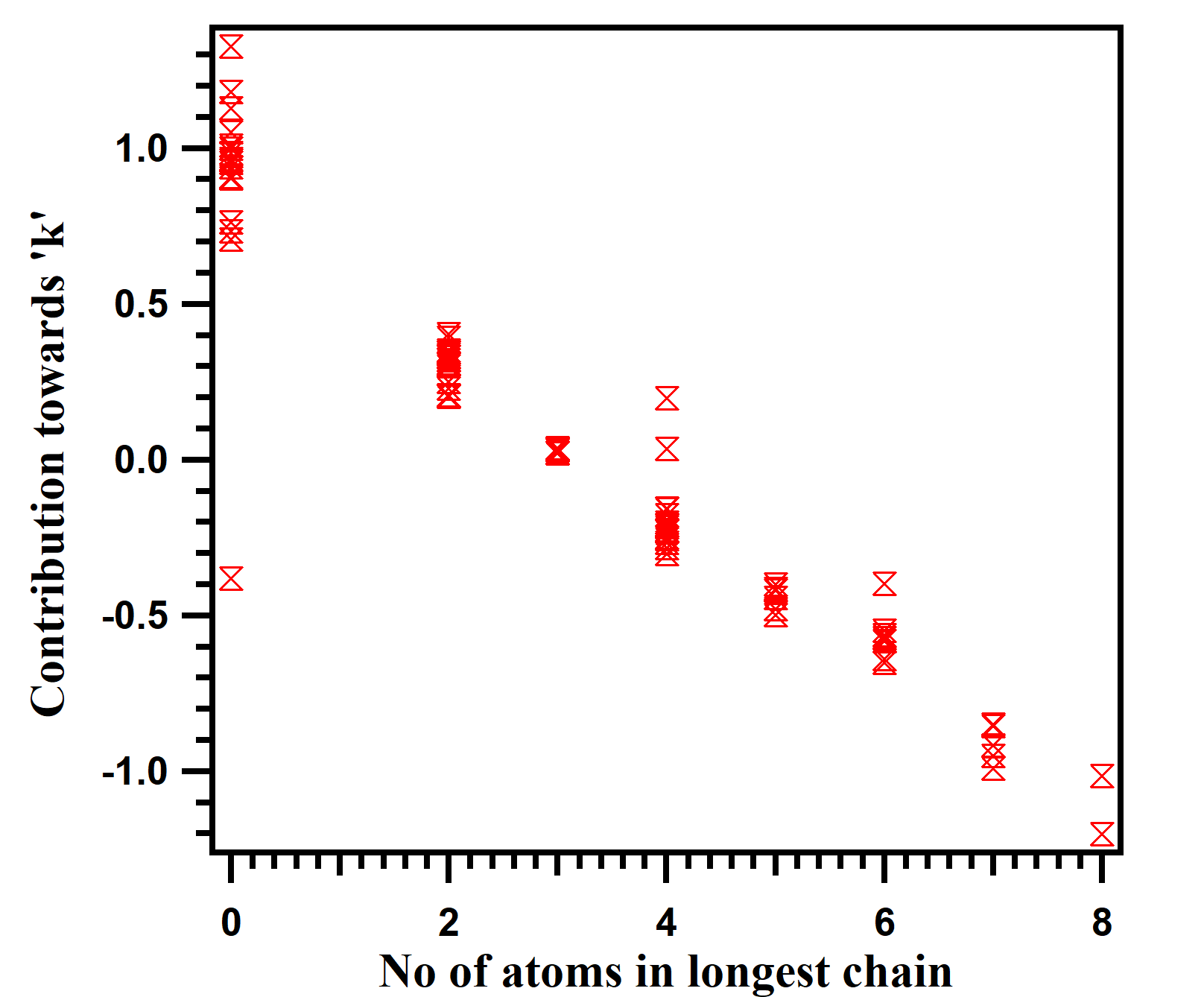}}
\end{minipage}%
\begin{minipage}{.6\linewidth}
\centering
\subfloat{\label{ap4:b}\includegraphics[width=\textwidth, height=6.5cm]{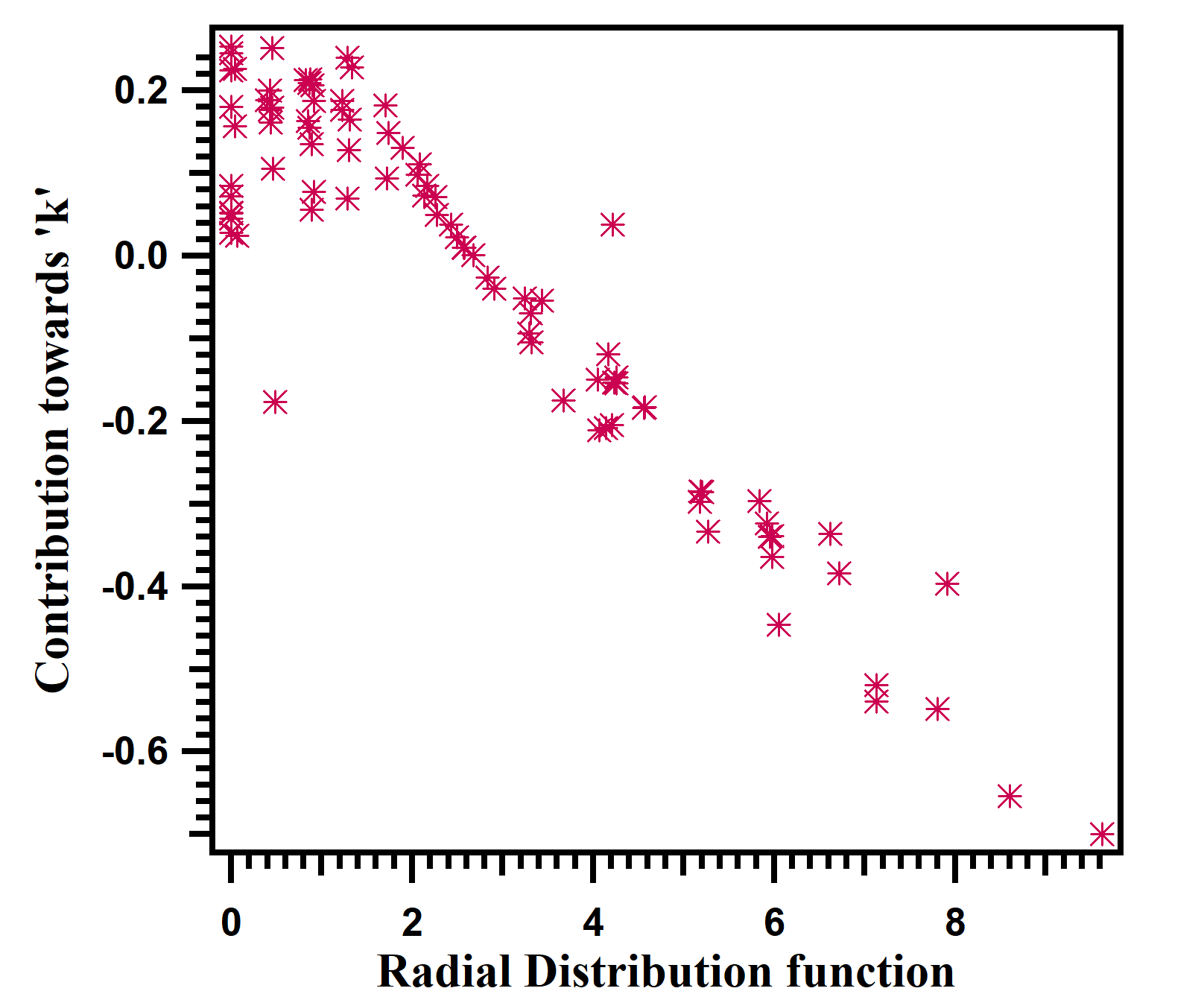}}
\end{minipage}\par\medskip
\label{fig:ap4}
\end{figure}

\begin{figure}[!ht]
\hspace*{-1.5cm}
\begin{minipage}{.6\linewidth}
\centering
\subfloat{\label{ap5:a}\includegraphics[width=\textwidth, height=6.5cm]{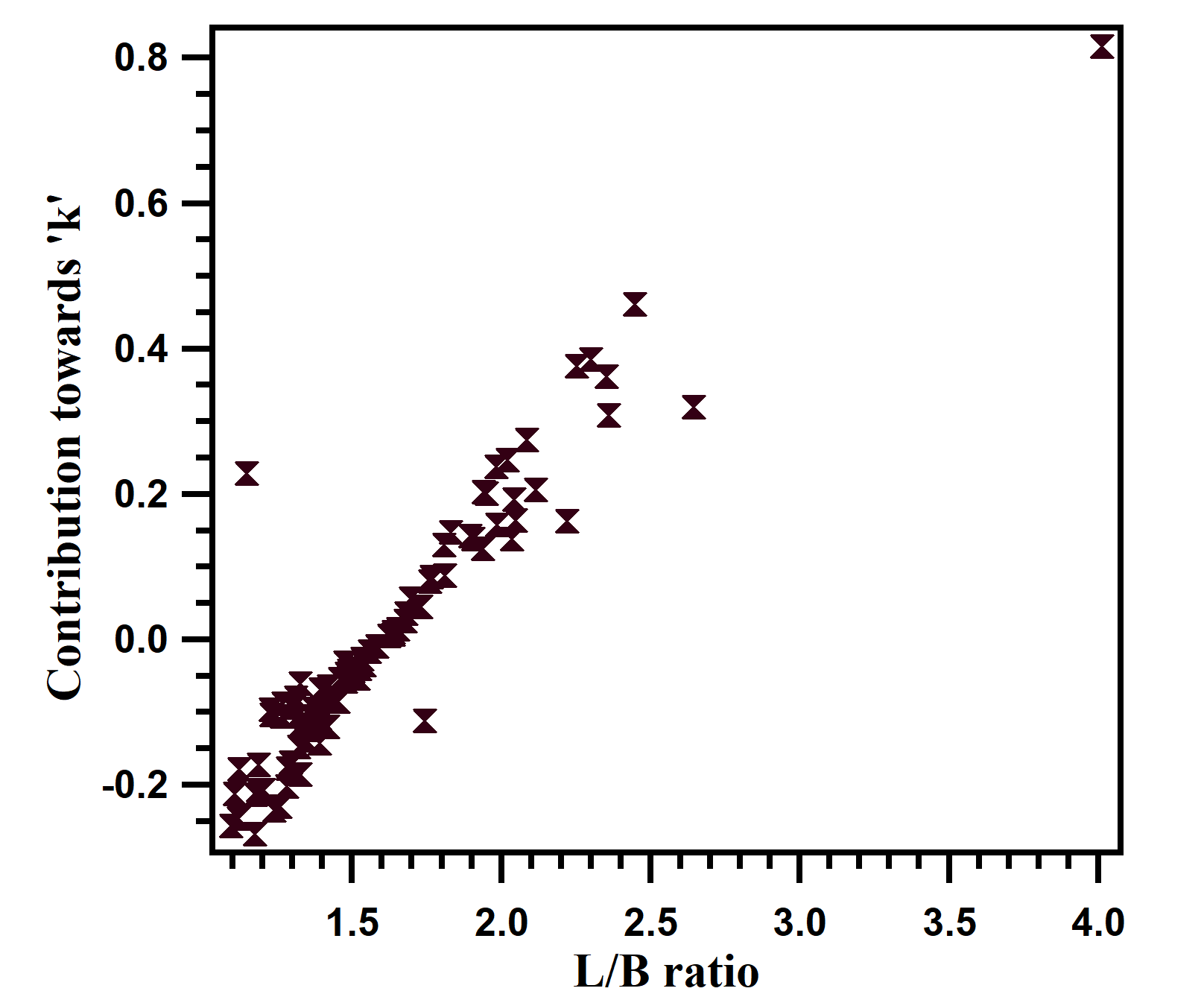}}
\end{minipage}%
\begin{minipage}{.6\linewidth}
\centering
\subfloat{\label{ap5:b}\includegraphics[width=\textwidth, height=6.5cm]{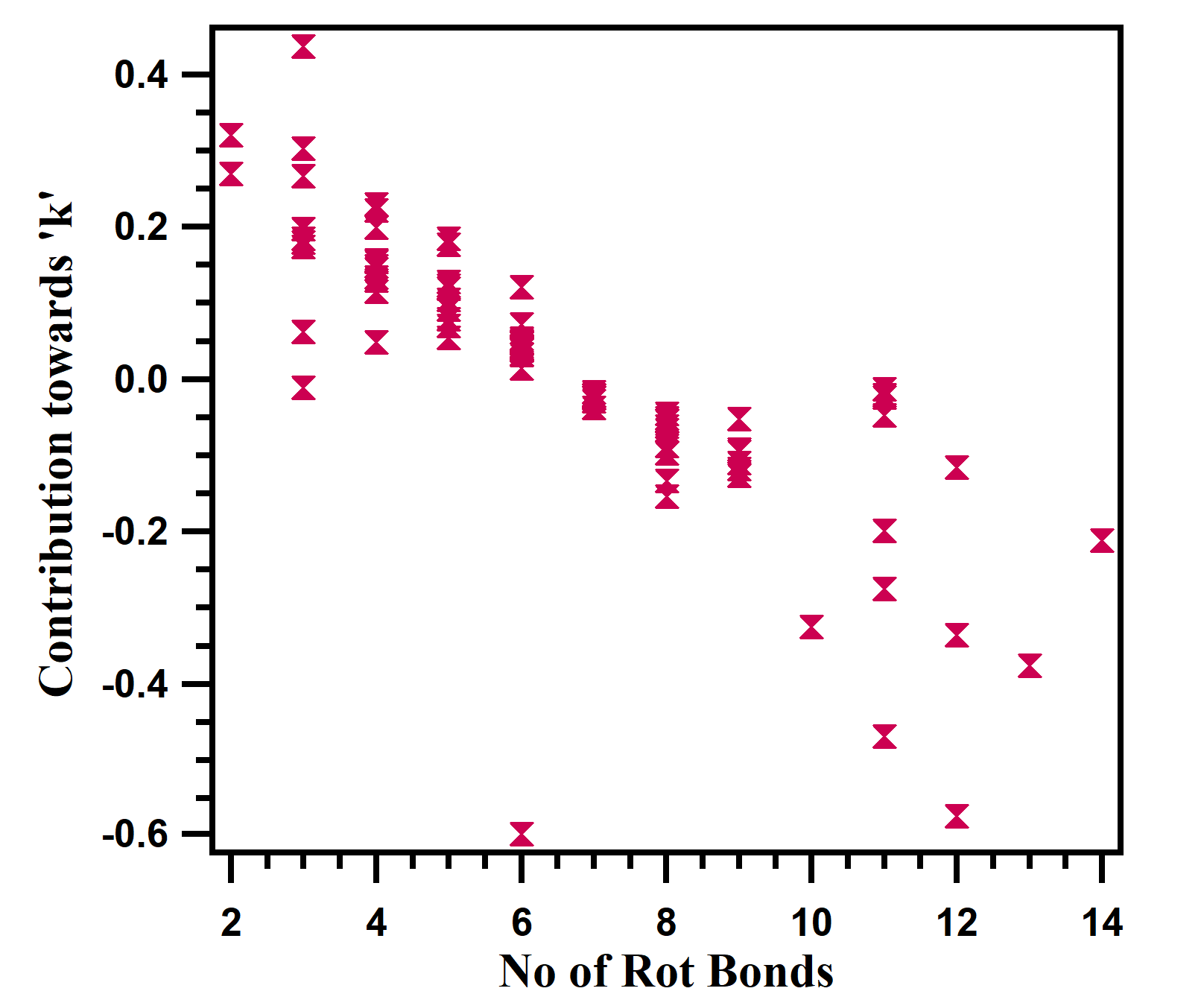}}
\end{minipage}\par\medskip
\label{fig:ap5}
\end{figure}

\end{document}